\documentclass[authoryear,review,3p,12pt,onecolumn]{elsarticle}
\usepackage[english]{babel}

\usepackage{setspace}

\usepackage{caption,blkarray}
\usepackage{bm}
\usepackage[labelformat=simple]{subcaption}

\usepackage{algorithm}
\usepackage{algpseudocode}

\usepackage{tikz}
\usepackage{array}
\newcolumntype{L}{>{\centering\arraybackslash}m{1cm}}
\usetikzlibrary{matrix,shapes,arrows,positioning,chains}
\usepackage[font=scriptsize, skip=0.2\baselineskip]{caption}
\usepackage{listings}
\usepackage{needspace}
\usepackage{rotating}
\usepackage{tabularx}
\usepackage{makecell}
\usepackage{booktabs}
\usepackage{geometry}
\usepackage{amsmath}
\usepackage{xcolor}
\usepackage{ragged2e}
\usepackage{float}
\usepackage{tablefootnote}

\usepackage{amsmath}
\usepackage{multirow}
\usepackage{epsfig, epstopdf}
\usepackage{colortbl}
\usepackage[backref=page]{hyperref}
\usepackage{lineno}
\usepackage{amssymb,amsmath,bm}
\usepackage{natbib}
\usepackage{setspace}
\usepackage{lscape}
\usepackage{longtable}
\usepackage{afterpage,latexsym,epsfig,float}
\bibpunct{(}{)}{;}{a}{}{,}
\RequirePackage{mathtools}
\allowdisplaybreaks
\usepackage[font=small,labelfont=bf]{caption}
\usepackage{hyperref}
\hypersetup{
     colorlinks   = true,
     citecolor    = gray
}

\definecolor{Gray}{gray}{0.9}

\begin{document}

\title{Why not a thin plate spline for spatial models? \\
A comparative study using Bayesian inference}

\author[Go]{Joaquin Cavieres}\corref{cor1}
\author[KAUST]{Paula Moraga}
\author[Wa]{Cole C. Monnahan}

\cortext[cor1]{Corresponding author. Email address: joaquin.cavieres@uni-goettingen.de}
\address[Go]{Universität Göttingen, Germany}
\address[KAUST]{Computer, Electrical and Mathematical Sciences and Engineering Division, King Abdullah University of Science and Technology (KAUST), Thuwal 23955-6900, Saudi Arabia}
\address[Wa]{Alaska Fisheries Science Center (NOAA), Seattle, WA, United States of America}

\begin{abstract}
\small{
Spatial modelling often uses Gaussian random fields to capture the stochastic nature of studied phenomena. However, this approach incurs significant computational burdens ($\mathcal{O}(n^3)$), primarily due to covariance matrix computations. Consequently, handling a large number of spatial data becomes computationally infeasible, particularly when non-Gaussian responses are modelled.

In this study, we propose to use a low-rank approximation of a thin plate spline as a spatial random effect in Bayesian spatial models. We compare its statistical performance and computational efficiency with the approximated Gaussian random field (by the SPDE method). In this case, the dense matrix of the thin plate spline is approximated using a truncated spectral decomposition, resulting in computational complexity of $\mathcal{O}(kn^2)$ operations, where $k$ is the number of knots. Bayesian inference is conducted via the Hamiltonian Monte Carlo algorithm of the probabilistic software Stan, which allows us to evaluate performance  and diagnostics for the proposed models. 

A simulation study reveals that both models accurately recover the parameters used to simulate data. However, models using a thin plate spline demonstrate superior execution time to achieve the convergence of chains compared to the models utilizing an approximated Gaussian random field. Furthermore, thin plate spline models exhibited better computational efficiency for simulated data coming from different spatial locations. 

In a real application, models using a thin plate spline as spatial random effect produced similar results in estimating a relative index of abundance for a benthic marine species when compared to models incorporating an approximated Gaussian random field. Although they were not the more computational efficient models, their simplicity in parametrization, execution time and predictive performance make them a valid alternative for spatial modelling under Bayesian inference.}

\end{abstract}

\begin{keyword}
Thin plate spline \sep approximated Gaussian random field \sep Bayesian modelling \sep spatial statistics \sep \texttt{tmbstan}
\end{keyword}

\maketitle

\section{Introduction}
\label{section1}

Spatial statistics methods have garnered significant interest in the scientific community in recent years. These methods allows analysts to incorporate spatial correlations between observations that would be challenging to consider using the traditional statistical methods. 

Particularly for data measured in a continuous spatial domain, one commonly employed methodology for incorporating uncertainty associated with a specific phenomenon is the use of a Gaussian random field (GRF). However, this approach can be computationally expensive, especially for a large number of observations. Due to this,  \cite{lindgren2011explicit} proposed an approximation of the GRF by a Gaussian Markov random field (GMRF,  \cite{rue2005gaussian}) through stochastic partial differential equations (SPDE method). With this approximation we can get computationally efficient estimation procedures by assuming a explicit spatial dependence for a spatial random field. This method was implemented using the integrated nested Laplace approximation for the estimation in a Bayesian hierarchical model in \citep{rue2009approximate,lindgren2015bayesian} and has been used in various research areas \citep{lindgren2022spde} due its fast computation and reliable results. 

On the contrary, a primary challenge in employing the SPDE method for Bayesian inference using MCMC methods is linked to the convergence of a hyperparameter associated with the Matérn correlation function.  A stationary Matérn field is represented by the parameters $\kappa$ and $\tau$ (the scale and variance, respectively), but in some cases, the latter faces convergence issues when using MCMC method (\cite{cavieres2021accounting}). An alternative to address this problem could involve using a different parameterization for the spatial random field. For instance, the parameterization proposed by \cite{lindgren2015bayesian} could be considered. 

On the other hand, a widely-used category of interpolants to approximate functions are the radial basis functions (RBFs, \cite{scheuerer2013interpolation}). RBFs, initially applied in machine learning problems, have found diverse applications in different fields \citep{wendland2004scattered, chiles2009geostatistics} and are a powerful tool for multivariate data interpolation in $n$ dimensions, as needed in spatial statistics \citep{cressie2022basis}. Several authors have introduced distinct methodologies involving RBFs for spatial modelling, especially using a thin plate splines (TPS) and smoothing thin plate splines (STPS). For instance, the work by \cite{wang2011smoothing} delves deeply into the realm of smoothing splines, \cite{wahba1975smoothing} shows how to choose the smoothing parameter for a periodic spline, \cite{hutchinson1985smoothing} introduces a procedure for calculating the trace of the influence matrix for a polynomial smoothing spline, while \cite{craven1978smoothing} estimated the appropriate degree of smoothing using generalized cross-validation (GCV). However, a disadvantage of using a TPS for spatial modelling is that, as the size of observations increases, the computational cost of estimation increases as well (\cite{cavieres2023thin}). This is due to the necessity of estimating a coefficient for each radial basis function associated with every observation. As a result,  computation becomes infeasible for a large scale data. To address this problem, \cite{wood2003thin} proposed an alternative which approximates the the splines matrix through truncated spectral decomposition. This approach facilitates efficient computation, even when dealing with a large number of spatial observations (\cite{wood2017generalized}). On the other hand, in the Bayesian context, \cite{wahba1983bayesian} and \cite{nychka1988bayesian} presented a method for estimating confidence intervals for smoothing splines. Furthermore, from the works of \cite{wahba1983bayesian}, \cite{kimeldorf1970correspondence}, and \cite{white2006bayesian}, the derivation of a spatial prior using a TPS provides a computationally efficient implementation without the need for MCMC methods for Bayesian computation. \cite{white2019direct} also proposed a method to incorporate a spatial prior based on TPS that can be sampled directly using Monte Carlo integration in a simple way. Lastly, \cite{cressie2022basis} conducted a comprehensive review of how RBFs can represent spatial stochastic processes and covariance functions, including RBFs for non-Gaussian response variables and non-Gaussian spatial processes. They also provide a review for using RBFs in spatial modelling within the R software (\cite{r2021}). However, none of these methodologies incorporate the option for Bayesian inference using an MCMC method (not approximated Bayesian inference as R-INLA does).


In this study, we compare two approaches for incorporating spatial dependence in Bayesian spatial models. The first spatial model uses an approximated GRF through the GMRF by the SPDE method, and a second spatial model introduces a TPS. The dense matrix of TPS is approximated following the proposal by \cite{wood2003thin} which allows for fast computation. We conduct a simulation study to assess the results, primarily focusing on the estimated parameters for each model, the posterior predictive distribution of the response variable, and computational efficiency (effective posterior samples generated per time). In a second study, we extended the model presented by \cite{cavieres2023thin} to a Bayesian framework. For all these analysis we use the Hamiltonian Monte Carlo (HMC) algorithm and its adaptive variant the no-U-turn sampler (NUTS, \cite{hoffman2014no}) of the probabilistic software Stan (\cite{gelman2015stan}, \cite{carpenter2017stan}) through  \texttt{Template Model Builder (TMB)} (\cite{kristensen2015tmb}) and its connection with \texttt{tmbstan} (\cite{monnahan2018no}).

The rest of the manuscript is organized as follows. In Section \ref{section2}, we present the methods used for the analyses, emphasizing the mathematical description of an approximated GRF and the TPS. Section \ref{section3} presents the metrics to evaluate and compare the models. 
In Section \ref{section4}, we describe the simulation study used to compare the two approaches for spatial modelling. Here, we present the parameters estimated, the computational efficiency of each model, and the posterior predictive performance for six different models based on different sizes of spatial locations. 
Section \ref{section5} shows a comparison of the models using a real case study, based on the estimation of an important vector in fisheries science.
We evaluate the predictive performance of each model by using approximate leave-one-one cross validation (loo-cv) and compare performance based on their computational efficiency. We conclude discussing the results and how our proposal can be a valid alternative for spatial models in the context of Bayesian inference. 

\section{Materials and methods}
\label{section2}

\subsection{Spatial random field}
\label{section2.1}

A continuous spatial random field in $d$ dimensions is defined by $\{Y(\boldsymbol{s}): \boldsymbol{s} \in D \subset \mathcal{R}^{d}\}$, where $\boldsymbol{s}$ is the location of process $Y(\boldsymbol{s})$ with variation in domain $D$. We say that the spatial random field is a Gaussian random field (GRF) if $\{Y(\boldsymbol{s}_{1}),....., Y(\boldsymbol{s}_{n})\} \sim \mathcal{N}_{n}(0, \Sigma)$, where the process is completely specified by mean function $\mu = \mathbb{E}(Y(\boldsymbol{s}))$ and covariance function $C(\boldsymbol{s}_{1}, \boldsymbol{s}_{2}) = Cov(Y(\boldsymbol{s}_{1}), Y(\boldsymbol{s}_{2}))$. The GRF can be assumed as stationary (strictly or weakly) and isotropic. A stationary isotropic random field has covariance functions that depend only on distance between points, that is $C(\boldsymbol{s}_{1}, \boldsymbol{s}_{2}) = Cov( ||\boldsymbol{s}_{1} - \boldsymbol{s}_{2}||)$. The dependence on the spatial structure is built through the covariance function, generally from the Matérn family which represents a very flexible class of functions that appear in many scientific disciplines \citep{stein2012interpolation,yuan2011models}. 


The approximated GRF uses a dependence conditional through a precision matrix $\boldsymbol{Q}$, which is the inverse of the covariance matrix $\boldsymbol{Q} = \boldsymbol{\Sigma}^{-1}$. \cite{lindgren2011explicit} proposed a new method to approximate a GRF with a GMRF through stochastic partial differential equations (SPDE). This methodology approximates a GMRF with weighted sum of simple basis functions, maintaining the domain's continuous space while the computational algorithms only see the discrete structures with Markov properties (\cite{lindgren2015bayesian}). With this stochastically-weak solution of the SPDE, the direct implication is to enable building a sparse precision matrix on a continuously-indexed region in a way that approximates a Matérn field, but lowers the computational cost from $\mathcal{O}(n^{2})$ in the GRF to $\mathcal{O}(n^{3/2})$ of the GMRF (\cite{blangiardo2015}). 

\subsubsection{The SPDE method}
\label{section2.1.1}

Consider the following expression for the the Matérn covariance function:


\begin{equation}\label{eq1}
\text{Cov}(\boldsymbol{s}_{i}, \boldsymbol{s}_{j}) =  \frac{\sigma^{2}}{2^{\nu-1}\Gamma(\nu)}(\kappa||\boldsymbol{s}_{i} - \boldsymbol{s}_{j}||)^{\nu}K_{\nu}(\kappa||\boldsymbol{s}_{i} - \boldsymbol{s}_{j}||),
\end{equation}

where $\Gamma$ is the Gamma function, $\kappa$ is the spatial scale parameter, $\sigma$ is the marginal standard deviation, $\nu$ is a smoothness parameter, and $K_{\nu}$ is the modified Bessel function of second kind. Considering the previous and the results presented by \cite{whittle1954stationary} and \cite{whittle1963stochastic}, the stationary solution of the SPDE method:

\begin{equation}
(\kappa^{2} - \Delta)^{\alpha/2}(\tau  u(\boldsymbol{s})) = \mathcal{W}(\boldsymbol{s}), \hspace{2mm} \boldsymbol{s} \in \mathbb{R}^{d}
\end{equation}
has a Matérn covariance function, with $\kappa > 0$, $\tau > 0$, $\alpha > d/2$, and where $\Delta$ is the Laplacian and $\mathcal{W}$ is the standard Gaussian spatial white noise. Using the finite element method (FEM), the spatial random field $u(\mathbf{s})$ can be represented through a set of basis functions:

\begin{equation}\label{eq3}
u(\boldsymbol{s}) = \sum^{n}_{k = 1} \psi_{k}(\boldsymbol{s})u_{k},
\end{equation}

with a sparse precision matrix $\boldsymbol{Q}$  constructed from the sparse matrices $\boldsymbol{C}$, $\boldsymbol{G}_{1}$ and $\boldsymbol{G}_{2}$. For a general case with $\alpha = 2$, the matrix $\boldsymbol{Q}$ has an explicit mathematical expression using the parameters $\tau$ and $\kappa$ in the covariance function such that:

\begin{equation}\label{eq4}
\boldsymbol{Q} = \tau^{2}(\kappa^{4}\boldsymbol{C} + 2\kappa^{2}\boldsymbol{G}_{1} + \boldsymbol{G}_{2}).
\end{equation}

Assigning a Gaussian distribution $u \sim \mathcal{N}(\boldsymbol{0}, \boldsymbol{Q}^{-1})$ yields continuous functions of $u(\boldsymbol{s})$, which at the same time are solutions of the SPDE method (\cite{lindgren2015bayesian}). \\

\subsection{Radial basis functions}
\label{section2.2}

The following definition has been adapted to the context of spatial statistics, however, Radial Basis Functions (RBFs) can also be used to approximate multivariate functions (\cite{buhmann2003radial}, \cite{wendland2004scattered}).  

Consider the spatial locations $\boldsymbol{s} \in \mathbb{R}^{d}$ and values $y_{i}$ corresponding to $\boldsymbol{s}_{i}$, $i = 1, \ldots, n$. Commonly, we assume that $y_{i}$ are obtained by sampling some (unknown) function $f$ at the sites $\{\boldsymbol{s}_{1}, \ldots, \boldsymbol{s}_{n}\}$ such that $y_{i} = f(\boldsymbol{s}_{i}), i = 1, \ldots, n$. Thus, we seek an approximation
$f: \mathbb{R}^{d} \rightarrow \mathbb{R}$ to the function $g: \mathbb{R}^{d} \rightarrow \mathbb{R}$ from which we assume the data were originated. If we are restricted to $D \subset \mathbb{R}^{d}$, and $D$ is prescribed (i.e., the spatial domain), then $f: D  \rightarrow \mathbb{R}$.

\subsubsection{Thin plate spline}
\label{section2.2.1}

A function $f$ is called a thin plate spline (TPS) on a set of spatial locations
$\{ \boldsymbol{s}_1,\dots,\boldsymbol{s}_n\}\subset\mathbb{R}^2$ if it has the following form  \citep{wahba1990spline, cavieres2023thin}:

\begin{equation}\label{eq6}
f(\boldsymbol{s}) = \sum^{n}_{i = 1}c_i\phi(||\boldsymbol{s} - \boldsymbol{s}_j||) +
\sum^3_{\ell = 1}d_{\ell}p_{\ell}(\boldsymbol{s}).
\end{equation}

Here, the TPS is defined in terms of a conditionally semidefinite RBFs \citep{wendland2004scattered} given by
$\boldsymbol{\Phi}(\boldsymbol{s}) = \|\boldsymbol{s}\|^{2} \log \|\boldsymbol{s}\|, \hspace{2mm} \boldsymbol{s} \in \mathbb{R}^{d},
$
where $\|\cdot\|$ is the Euclidean distance. The basis $\{p_1, \ldots, p_l\}$ belongs to:
\begin{align*}
    \mathcal{P}^2_1 = \{ p:\mathbb{R}^2 \rightarrow \mathbb{R} \mid \exists a,b,c\in\mathbb{R}:\;
    p(\boldsymbol{s}) = a \boldsymbol{s}_1 + b\boldsymbol{s}_2 + c \}.
\end{align*}

If the vector of coefficients $\boldsymbol{c} = (c_1,\dots, c_n)^\top$ satisfies $\sum_{i=1}^n c_i p (\boldsymbol{s}_i) = 0$ for all $p\in\mathcal{P}^2_1$, then $f$ is called a natural TPS. It is sufficient to impose the last identity only on a basis of $\mathcal{P}^2_1$, which leads to $\boldsymbol{P}^\top\boldsymbol{c} = \boldsymbol{0}$ for the matrix $\boldsymbol{P}\in\mathbb{R}^{n\times 3}$ given by $\boldsymbol{P}_{i,j} = p_j(\boldsymbol{s}_i)$.

\subsubsection{Smoothing thin plate spline}
\label{section2.2.2}

Let $f$ be an approximating function for $n$ observations $y_{i}$ at spatial locations $\boldsymbol{s}_{i}$. Hence, we can propose a spatial statistical model as:

\begin{equation}\label{eq7}
y_{i} = f(\boldsymbol{s}_{i}) + \epsilon_{i},\quad i=1,\dots, n,
\end{equation}
where $\epsilon_{i}$ is a random error assumed as $\mathcal{N}(0,1)$. The function $f$ can be estimated by minimizing
the residual sum of penalized least squares terms such that:

\begin{equation}\label{eq8}
S(f) = \sum^{n}_{i = 1} |y_{i} - f(\boldsymbol{s}_{i})|^{2} + \lambda J(f).
\end{equation}

Here, the component $J(f)$ represents a penalty function related to the smoothness of $f$, and $\lambda$ serves as a penalty parameter regulating the trade-off between data fit and smoothness. For the TPS in $\mathbb{R}^2$, the penalty function takes the form:

\begin{equation}\label{eq9}
J(f) = \int_{\mathbb{R}^{2}} \bigg (\frac{\partial^2f}{\partial \boldsymbol{s}_1^2}\bigg )^{2}(\boldsymbol{s})
+2\bigg (\frac{\partial^2f}{\partial \boldsymbol{s}_1\partial\boldsymbol{s}_2}\bigg )^{2}(\boldsymbol{s})
+\bigg (\frac{\partial^2f}{\partial \boldsymbol{s}_2^2}\bigg )^{2}(\boldsymbol{s})\,d\boldsymbol{s},
\end{equation}

Following \citep{green1993nonparametric}, it has a closed form representation as:

\begin{equation*}
    J(f) = \boldsymbol{c}^\top \boldsymbol{E}\boldsymbol{c}.
\end{equation*}

The objective is to minimize Equation \eqref{eq8} over the set of natural TPS. To that end,
we define the matrix $\boldsymbol{E}\in\mathbb{R}^{n\times n}$ by $\boldsymbol{E}_{j,i} = \phi(\| \boldsymbol{s}_j-\boldsymbol{s}_i\|_2)$.  Hence, the minimization of $S(f)$ defined in Equation \eqref{eq8} for the TPS can be expressed as:

\begin{equation}\label{eq10}
    \min_{\substack{\boldsymbol{c}\in\mathbb{R}^n,\boldsymbol{d}\in\mathbb{R}^3\\ \boldsymbol{P}^\top \boldsymbol{c}=\boldsymbol{0}}}
    ||\boldsymbol{y} - \boldsymbol{E}\boldsymbol{c} - \boldsymbol{P}\boldsymbol{d}||_2^{2} + \lambda\boldsymbol{c}^{T}\boldsymbol{E}\boldsymbol{c},
\end{equation}

where the matrix $\boldsymbol{E}$ is replaced by a truncated spectral decomposition $\boldsymbol{E}_{k} = \boldsymbol{U}_{k}\boldsymbol{D}_{k}\boldsymbol{U}^{T}_{k}$ (\cite{wood2003thin}), $\boldsymbol{P}$ was already defined and $\boldsymbol{d} = (d_{1}, \ldots, d_{3})^{T}$ from Equation (\ref{eq6}).

\section{Performance metrics}
\label{section3}

Having defined the models, we will introduce different metrics to evaluate and compare them. Since that our main goal is to conduct a fair comparison between models using an approximated GRF and a TPS as spatial random effects, we will consider the quantities reported by Stan based on its Markov Chain Monte Carlo (MCMC) algorithm. The metrics and quantities are:

\begin{enumerate}
    \item \underline{Execution time}: total computational time of all chains, excluding model compilation.  
    \item \underline{Effective sample size (ESS)}:  in simple terms, ESS refers to the estimated number of independent draws sampled from the posterior distribution for each parameter (\cite{aki2020rank}). 
    \item \underline{Computational efficiency}: the number of effective samples generated per time, calculated as min (ESS)/ run time.
    \item \underline{Rhat ($\hat{R}$ statistic)}: measures the degree of mixing of the chains and its convergence with values $\hat{R}>1.05$ indicating evidence of lack of convergence (\cite{aki2020rank}). 
    \item \underline{Posterior predictive distribution (ppd)}: it is a way to validate a Bayesian model by ensuring that the observed data could have realistically come from the model (\cite{gelman2014bayesian}). Represents the distribution of unobserved or future data points given the observed data and the estimated parameters of the model. 
\end{enumerate}

\section{Simulation study}
\label{section4}
 
In this section, we assess the performance of the TPS as spatial random effect compared with the approximated GRF ($\approx$GRF) using a simulation study.
Specifically, we simulate data from a set of spatial locations and then fit the data using both approaches and compare parameter estimates relative to those used in the simulation, computational efficiency, and the posterior predictive performance of each model.

\subsection{Model specification}

Specifically, we simulate data as
\begin{equation}\label{eq11}
    y_{i} = \beta_{0} + \beta_{1}x_{1i} + f(\boldsymbol{s}_{i}) + \epsilon_{i}, \hspace{2mm} i = 1, \ldots, n,
\end{equation}
where $y_{i}$ is the response variable, $\beta_{0}$ is the intercept and $\beta_{1}$ is the parameter associated with a fixed covariate $x_{1}$ simulated from an uniform distribution, $x_{1} \sim \mathcal{U}(0, 1)$. $f(\boldsymbol{s}_{i})$ is the spatial random effect ($\approx$GRF or TPS) and $\epsilon_{i}$ is the $\textit{i.i.d}$ random error, $\epsilon_i \sim \mathcal{N}(0, \sigma)$. 

To evaluate the performance of both models regarding the spatial random effect, we will consider different sizes of spatial locations (SL). The SL were built from 100 to 600 spatial random points, hence there are 6 different sizes of spatial locations; SL1, SL2, SL3, SL4, SL5 and SL6. The values for the parameters in the simulation are: $\beta_{0} = 1.0$, $\beta_{1} = 2.0$, $\sigma = 0.1$.
For convenience, we refer to the model incorporating a $\approx$GRF as \textbf{M-GRF}, and the model including a TPS as as \textbf{M-TPS}. Next, we give the mathematical formulation for both models:

\begin{enumerate}
\setlength\itemsep{1.5em}
    \item \textbf{M-GRF} adopts the classical structure of a hierarchical model, that is:
    \begin{align}
\boldsymbol{\theta} \hspace{2mm} \sim & \hspace{2mm} \pi(\boldsymbol{\theta}) \nonumber\\
\boldsymbol{u} \mid \boldsymbol{\theta} \hspace{2mm} \sim & \hspace{2mm} \mathcal{N}(\boldsymbol{0}, \boldsymbol{Q}(\boldsymbol{\theta})^{-1}) \nonumber\\
\boldsymbol{y} \mid \boldsymbol{u}, \boldsymbol{\theta} \hspace{2mm} \sim & \hspace{2mm} \prod_{i}\pi(y_{i} \mid \eta_{i}, \boldsymbol{\theta}) \nonumber,
\end{align}

where $\boldsymbol{\theta}$ is the vector of hyperparameters $\{\text{log}(\tau), \text{log}(\kappa)\}$, $\boldsymbol{u}$ is a Gaussian random field ($\approx$GRF), $\boldsymbol{Q}(\boldsymbol{\theta})$ is the precision matrix, $\boldsymbol{\eta} = \boldsymbol{X}\boldsymbol{\beta} + \mathbf{A}\boldsymbol{u}$ where $\boldsymbol{X}$ is the model matrix considering $[\boldsymbol{1}, \boldsymbol{x}^{T}]$, $\boldsymbol{\beta} =(\beta_{0}, \beta_{1})$ are the fixed parameters in the model,  $\mathbf{A}$ is the projector matrix and $\eta_{i}$ is associated to the observations $y_{i}$. Finally, $\boldsymbol{y}$ is the vector of the response variable $f(\cdot \mid \boldsymbol{u}, \boldsymbol{\theta})$, commonly belonging from the exponential family of distributions. 
The prior distribution for the parameters is as follows; for the fixed effects we assume that $\beta_{0} \sim \mathcal{N}(1, 2)$, $\beta_{1} \sim \mathcal{N}(1, 2)$ and $\sigma \sim \text{Cauchy}(0, 5)$. For the hyperparameters of the $\approx$GRF we assume $\tau \sim \mathcal{N}(0, 0.5)$ and $\kappa \sim \mathcal{N}(0, 0.5)$.

\item \textbf{M-TPS} considers the expression in \ref{eq11} as follow (in matrix form) :
\begin{eqnarray}\label{eq13}
\boldsymbol{y} = \boldsymbol{X} \boldsymbol{\beta} + \boldsymbol{E}_{k} \boldsymbol{c} + \boldsymbol{\epsilon},
\end{eqnarray}

where $\boldsymbol{y}=(y_{1},\ldots,y_{n})^{\top}$, $\boldsymbol{X}$ and $\boldsymbol{\beta}$ as described before, $\boldsymbol{E}_{k}$ is the splines matrix, $\boldsymbol{c}$ are the spline coefficients, and $\boldsymbol{\epsilon}=(\varepsilon_{1},\ldots,\varepsilon_{n})^{\top}$ is as before. In this case, the term $\boldsymbol{P}\boldsymbol{d}$ of the expression in \ref{eq10} is removed to avoid identifiability problems. 
Following to \cite{cavieres2023thin}, the penalized log-likelihood of $\boldsymbol{\theta} = (\boldsymbol{\beta}^{T}, \boldsymbol{c}^{T}, \sigma, \lambda)$ can be expressed as:

\begin{equation}\label{eq14}
\ell_p(\boldsymbol{\theta}, \lambda) =
\ell(\boldsymbol{\theta}) - \frac{\lambda}{2}
\boldsymbol{c}^{\top} \boldsymbol{S}\boldsymbol{c} \hspace{0.1cm},
\end{equation}

where $\ell(\boldsymbol{\theta})$ is the log-likelihood for the Gaussian density function, $\lambda$ is the penalty parameter and $\boldsymbol{S}$ is a rank deficient matrix. The prior distribution for coefficients of the TPS are assumed as $\boldsymbol{c} \sim \mathcal{N}(0, 1)$.
\end{enumerate}


All the analysis were done using a Dell laptop with an 11th Gen Intel(R) Core(TM) i7-1165G7 @ 2.80GHz 1.69 GHz processor and 16.0 GB (15.4 GB usable) of RAM. For the comparison, both models were run using 4 chains with 700 warm up samples (per chain), 4500 total iterations, $\texttt{adap\_delta} = 0.95$ and $\texttt{max\_treedepth= 13}$ (controlling the target acceptance rate and maximum trajectory length of each iteration, respectively). We report computational efficiency as the minimum number of effective samples, as calculated by the R package ``rstan" (\cite{rstan_cite}),  divided by the total run time in seconds (i.e., how many effective samples are produced per second). For the case of \textbf{M-TPS}, the estimation of the penalty parameter will not be presented in the results. 

\subsection{Simulation results}
\label{sec:simulation-results}

Table \ref{table:table1} shows the mean, $25\%$ and $75\%$ percentiles of the posterior distribution for the parameters, relative error and the execution time for all the models. In each model, the parameters estimated by $\textbf{M-GRF}$ and $\textbf{M-TPS}$ are similar to the ``real"  parameters used in the simulation. In SL1 and SL3 the parameters estimated by \textbf{M-GRF} for $\beta_{0}$ and $\beta_{1}$ were closer to the true values, while in SL4 and SL5 these same parameters were estimated much better by \textbf{M-TPS}.    


\newpage
\begin{table}[!htbp]
\setlength{\tabcolsep}{4pt}
\caption{Means, $25\%$, and $75\%$ percentiles of the posterior distributions, relative error and execution time for \textbf{M-GRF} and \textbf{M-TPS} in the different simulated spatial locations (SL).}
\centering
\scalebox{0.95}{
\begin{tabular}{cccccccccccc}
\hline
    &  & \multicolumn{5}{c}{\textbf{M-GRF}} &   \multicolumn{5}{c}{\textbf{M-TPS}}\\
   \hline
 SL   & Real values & Mean & 25\% & 75\% & R.error & E.time     & Mean  & 25\% & 75\% & R.error & E.time \\
 \hline
\multirow{3}{*}{SL1} & $\beta_{0} = 1.0$ & 1.01 & 0.96 & 1.05 & 0.01 &  \multirow{3}{*}{\textcolor{red}{$\sim$ 6 min}} & 1.03 & 1.01 & 1.05 & 0.03 & \multirow{3}{*}{\textcolor{blue}{$\sim$ 2 min}} \\
                     & $\beta_{1} = 2.0$ & 2.05 & 2.00 & 2.09 & 0.05 &  & 1.97 & 1.94 & 2.00 & 0.03\\
                     & $\sigma    = 0.1$ & 0.10 & 0.09 & 0.11 & 0.00 &  & 0.11 & 0.11 & 0.12 & 0.01\\
\hline
\multirow{3}{*}{SL2}& $\beta_{0}   = 1.0$ & 1.02  & 0.98 & 1.05 & 0.02 & \multirow{3}{*}{\textcolor{red}{$\sim$ 10 min}} & 1.00 & 0.99 & 1.01 & 0.00 &\multirow{3}{*}{\textcolor{blue}{$\sim$ 3 min}} \\
                      & $\beta_{1} = 2.0$ & 1.99  & 1.97 & 2.01 & 0.01 & & 2.00 & 1.98 & 2.02 & 0.00\\
                      & $\sigma    = 0.1$ & 0.09  & 0.09 & 0.10 & 0.01 & & 0.11 & 0.11 & 0.12 & 0.01\\
\hline
\multirow{3}{*}{SL3} & $\beta_{0}  = 1.0$ & 0.98 & 0.94 & 1.03 & 0.02 & \multirow{3}{*}{\textcolor{red}{$\sim$ 18 min}} & 0.98 &	0.97 & 0.99  & 0.02 & \multirow{3}{*}{\textcolor{blue}{$\sim$ 4 min}} \\
                      & $\beta_{1} = 2.0$ & 2.00 & 1.98 & 2.01 & 0.00 & &  2.03 & 2.02 & 2.05 & 0.03\\
                      & $\sigma    = 0.1$ & 0.09 & 0.09 & 0.10 & 0.01 & &  0.10 & 0.10 & 0.10 & 0.00\\
\hline
\multirow{3}{*}{SL4} & $\beta_{0}  = 1.0$ & 1.02 & 0.98 & 1.05 & 0.02 & \multirow{3}{*}{\textcolor{red}{$\sim$ 20 min}} & 1.00 &	0.99 & 1.01  & 0.00& \multirow{3}{*}{\textcolor{blue}{$\sim$ 7 min}} \\
                      & $\beta_{1} = 2.0$ & 1.97 & 1.96 & 1.99 & 0.03 & & 1.99 & 1.98 & 2.00 & 0.01\\
                      & $\sigma    = 0.1$ & 0.10 & 0.09 & 0.10 & 0.00 & & 0.10 & 0.09 & 0.10 & 0.00\\
\hline
\multirow{3}{*}{SL5} & $\beta_{0}  = 1.0$ & 0.93 & 0.90 & 0.96 & 0.07 & \multirow{3}{*}{\textcolor{red}{$\sim$ 18 min}} & 1.00 &	0.99 & 1.01  & 0.00 & \multirow{3}{*}{\textcolor{blue}{$\sim$ 14 min}} \\                       & $\beta_{1}  = 2.0$ & 2.01 & 2.00 & 2.02 & 0.01 & & 1.99 & 1.98 & 2.00 & 0.00\\
                     & $\sigma     = 0.1$ & 0.09 & 0.09 & 0.10 & 0.01 & & 0.10 & 0.09 & 0.10 & 0.00\\
\hline
\multirow{3}{*}{SL6} & $\beta_{0}  = 1.0$ & 0.96 & 0.93 & 0.99 & 0.04 & \multirow{3}{*}{\textcolor{red}{$\sim$ 26 min}} & 1.02 &	1.01 & 1.02  &  0.02& \multirow{3}{*}{\textcolor{blue}{$\sim$ 18 min}} \\                       & $\beta_{1}  = 2.0$ & 2.01 & 2.00 & 2.02 & 0.01 & & 1.98 & 1.97 & 1.99 & 0.02\\
                     & $\sigma     = 0.1$ & 0.10 & 0.10 & 0.10 & 0.00 & & 0.10 & 0.09 & 0.10 & 0.00\\
\hline
\end{tabular}}
\label{table:table1}
\end{table}

\footnotesize{R.error = $||\text{Real value} - \text{Mean}||$
\vspace{0.2cm}

E.time = Execution time.}
\normalsize

\newpage

MCMC algorithms generate samples from the posterior distribution, and the quality of the exploration it performs within this distribution is an important factor in assessing its effectiveness. Here we used the NUTS algorithm for both models, so there is no problem in comparing the computational efficiency between the models. Considering this, the computational efficiency for $\textbf{M-TPS}$ is greater than $\textbf{M-GRF}$ for every model. Furthermore, it is essential to highlight the difficulty in parameterizing  $\textbf{M-GRF}$,
as the hyperparameter $\tau$ often exhibits convergence issues across chains during the analyses. 

\begin{figure}[hbt!]
\centering
\includegraphics[width=12cm, height=8cm]{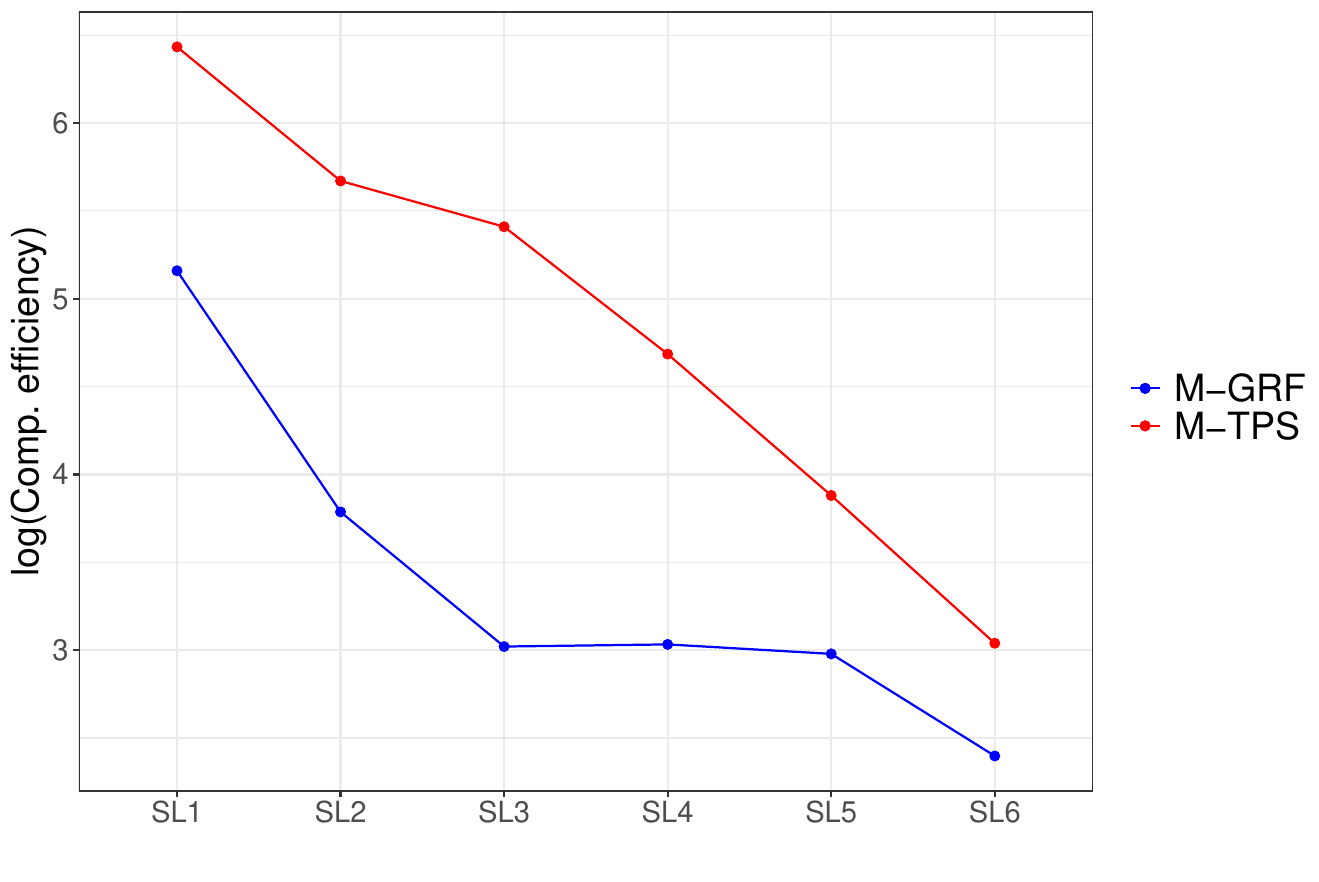}
 \caption{Computational efficiency in log scale (log(\texttt{min (ESS)/ time})) calculated for different spatial locations (SL) in \textbf{M-GRF} and \textbf{M-TPS}.}\label{fig:fig1}
\end{figure}

Figure \ref{fig:fig1} shows that the $\textbf{M-TPS}$ exhibits better computational efficiency (in log scale) compared to the \textbf{M-GRF} in all the SL. Although the graph suggests that further increasing the number of spatial locations (SL) could eventually represents a similar computational efficiency for \textbf{M-GRF} and \textbf{M-TPS}, this was done intentionally to try to make a fair comparison between models, that is to say, as we increased the number of SL,  the number of knots in the \textbf{M-TPS} and the number of triangles of the mesh for the \textbf{M-GRF} also increased. However, if we consider a small number of knots to obtain the truncated basis functions for the \textbf{M-TPS}, it is still computationally efficient and accurately estimates the parameters $\beta_{0}$, $\beta_{1}$ and $\sigma$ used in the simulated data (\ref{appendix:appendix1}). 

\begin{figure}[hbt!]
\centering
\includegraphics[width=14cm, height=6.8cm]{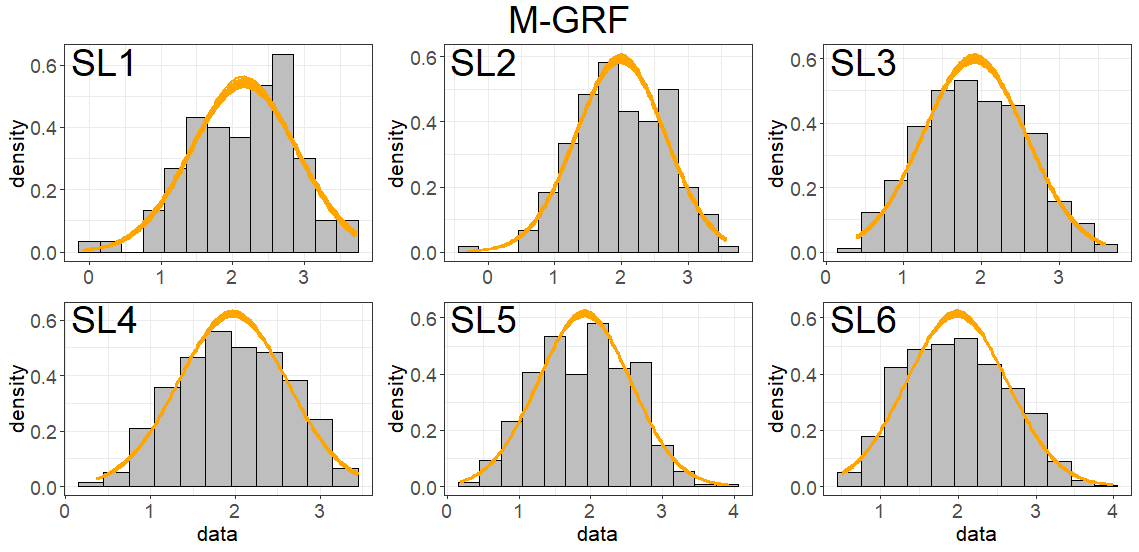}
 \caption{Simulated response variable ``data" and 100 posterior predictive distributions ``y\_ppd" from the \textbf{M-GRF} model for different SL.}\label{fig:fig2}
\end{figure}
\begin{figure}[hbt!]
\centering
\includegraphics[width=14cm, height=6.8cm]{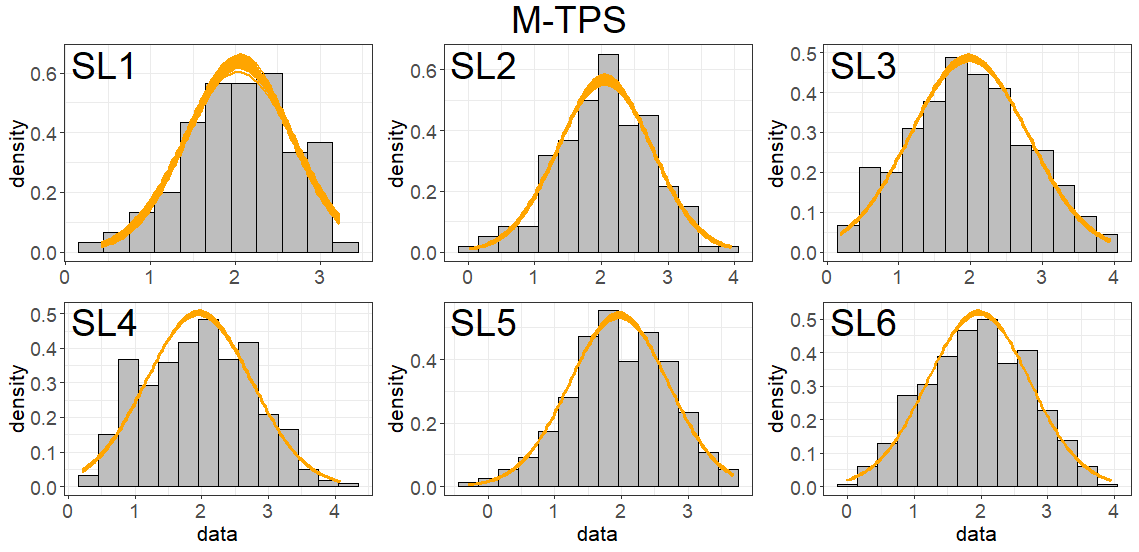}
\caption{Simulated response variable ``data" and 100 posterior predictive distributions ``y\_ppd" from the \textbf{M-TPS} model for different SL.}\label{fig:fig3}
\end{figure}

\newpage
To evaluate the predictive performance of the response variable (``data") from the parameters estimated by \textbf{M-GRF} and \textbf{M-TPS}, we computed the posterior predictive distribution (``y\_ppd") in each SL. As we can see in Figures \ref{fig:fig2} and  \ref{fig:fig3}, both models appear to produce simulated data that closely match the observed data.  

\newpage
The mean values for the posterior predictive distribution (``y\_ppd") are quite similar in both models. For example, if we consider the mean value of the simulated response (``data") in SL3 for \textbf{M-GRF}, the mean value of ``y\_ppd" is 1.91 and the posterior predictive mean is 1.91. In the same SL3, but now for \textbf{M-TPS}, the mean value of ``data" is 1.98, with the same mean value for ``y\_ppd". The standard deviations of the simulated data also are similar between the models but they are higher in the \textbf{M-TPS}. The difference between 0.63 and 1.03 is 0.4 for \textbf{M-TPS}, while for \textbf{M-GRF} the difference of those values is 0.1 (Table \ref{table:table2}).

\begin{table}[hbt!]
\setlength{\tabcolsep}{10pt}
\caption{Mean and standard deviations (Sd) for ``data" and ``y\_ppd" from \textbf{M-GRF} and \textbf{M-TPS} for 100 replicated data sets.}
\label{table:table2}\textbf{}
\centering
\scalebox{1.0}{
\begin{tabular}{LLLLLLLLL}
\hline
 & \multicolumn{4}{c}{\textbf{M-GRF}} & \multicolumn{4}{c}{\textbf{M-TPS}} \\
\hline
SL     & Mean data & Sd data & Mean y\_ppd & Sd y\_ppd & Mean data & Sd data & Mean y\_ppd  & Sd y\_ppd  \\
\hline
SL1    &    2.16   &  0.74  &    2.16     &    0.74   &    2.04     & 0.62  &        2.04        & 0.63     \\
SL2    &    1.99   &  0.68  &    1.99     &    0.67   &    2.05     & 0.70  &        2.05        & 0.71     \\
SL3    &    1.91   &  0.68  &    1.91     &    0.66   &    1.98     & 0.82  &        1.98        & 0.82    \\
SL4    &    1.97   &  0.65  &    1.97     &    0.64   &    1.97     & 0.80  &        1.96        & 0.97      \\
SL5    &    1.91   &  0.65  &    1.91     &    0.65   &    1.96     & 0.74  &        1.96        & 0.81     \\
SL6    &    1.98   &  0.66  &    1.98     &    0.65   &    1.97     & 0.77  &        1.97        & 1.03      \\
\hline
\end{tabular}}
\end{table}

\newpage
\section{Real case study: Index of relative abundance of sea urchin in Chile.}
\label{section5}

In this section we compare four models which estimate a standardized index of relative of abundance of marine resources that is commonly used in stock assessments models to provide information on trends in population abundance (\cite{hilborn2013quantitative, ney1993practical}). The models are typically fit to commercial catch-per-unit-effort (CPUE) data, and the models attempt to control for factors that affect changes in CPUE other than changes in the underlying population size (\cite{maunder2004standardizing}). We fit the models to commercial CPUE data from the sea urchin (\textit{Loxechinus albus}) in the southern region of Chile.   

\subsection{Sea urchin data in Chile}

\begin{figure}[h!]
\centering
\includegraphics[width=14cm, height=10cm]{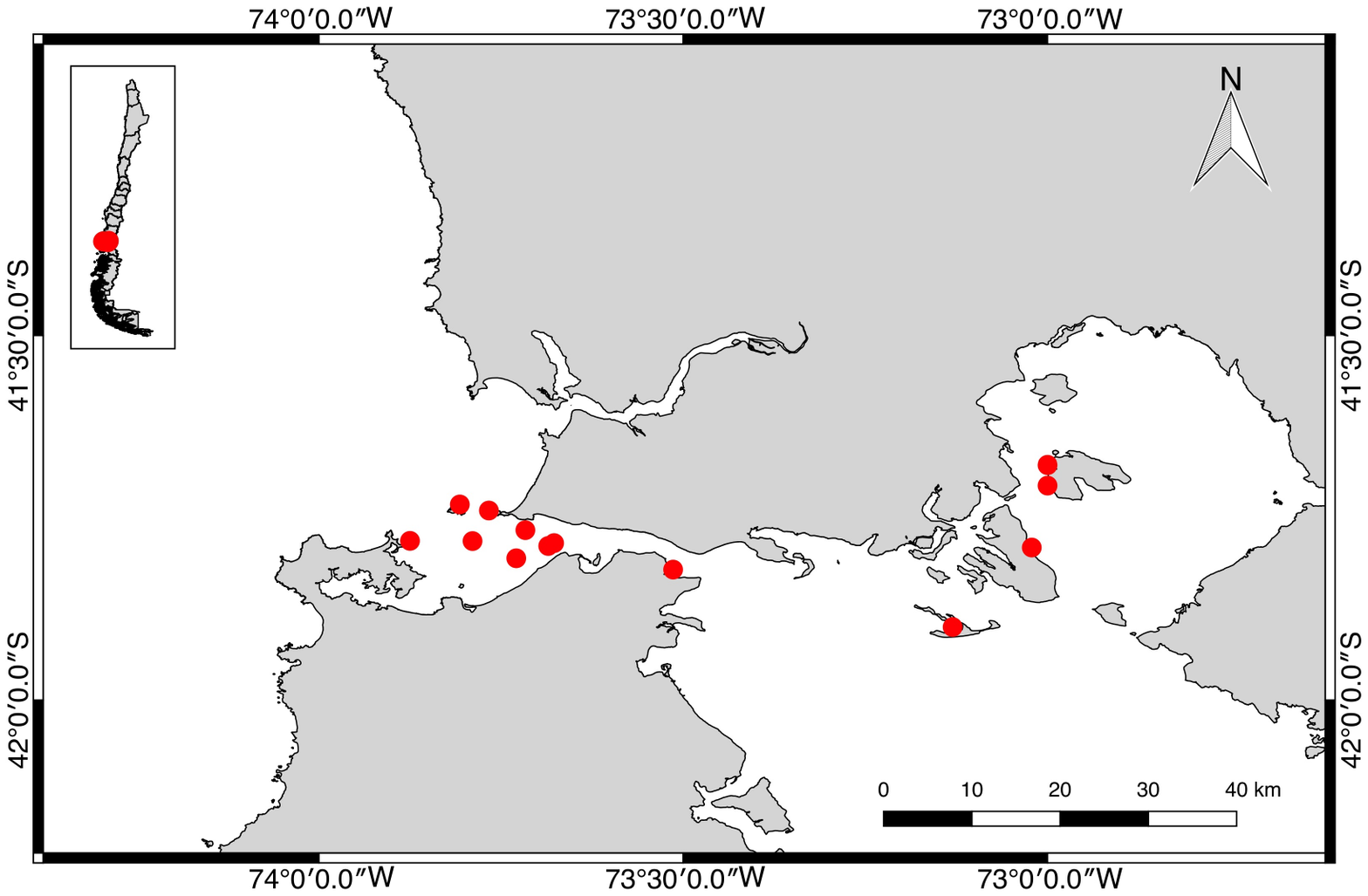}
\caption{Locations of fishing sites (red points) of sea urchin in the north of Isla de Chilóe Chile. }\label{fig:fig4}
\end{figure}

Due to their large-scale spatial metapopulation structure, sea urchin subpopulations are interconnected by larval dispersion, so the recovery of local abundance depends on the distance, and hydrodynamic characteristics of their spatial domain. The data we analyze contains 3,148 observations, where the number of observations also varies among the fishing sites (Figure \ref{fig:fig4}).  The relative index of abundance is estimated using the response variable ``CPUE", which in turn, is explained by a set of covariates: a spatial random effect representing 13 sites of fishing, a temporal variable called ``year'' (From 1996 - 2016), the ``depth'' (average depth of catches), ``season'' (season of the year, 1, 2, 3) and ``destine'' (destine of the sales target, 1 or 2). The relative index is defined as the coefficients estimated for the variable ``year'' and declared as factor in the model. For a more detailed explanation of the data refer to \cite{cavieres2021accounting}. 

\subsection{Models}

\begin{figure}[h!]
\centering
\includegraphics[width=10.5cm, height=6.5cm]{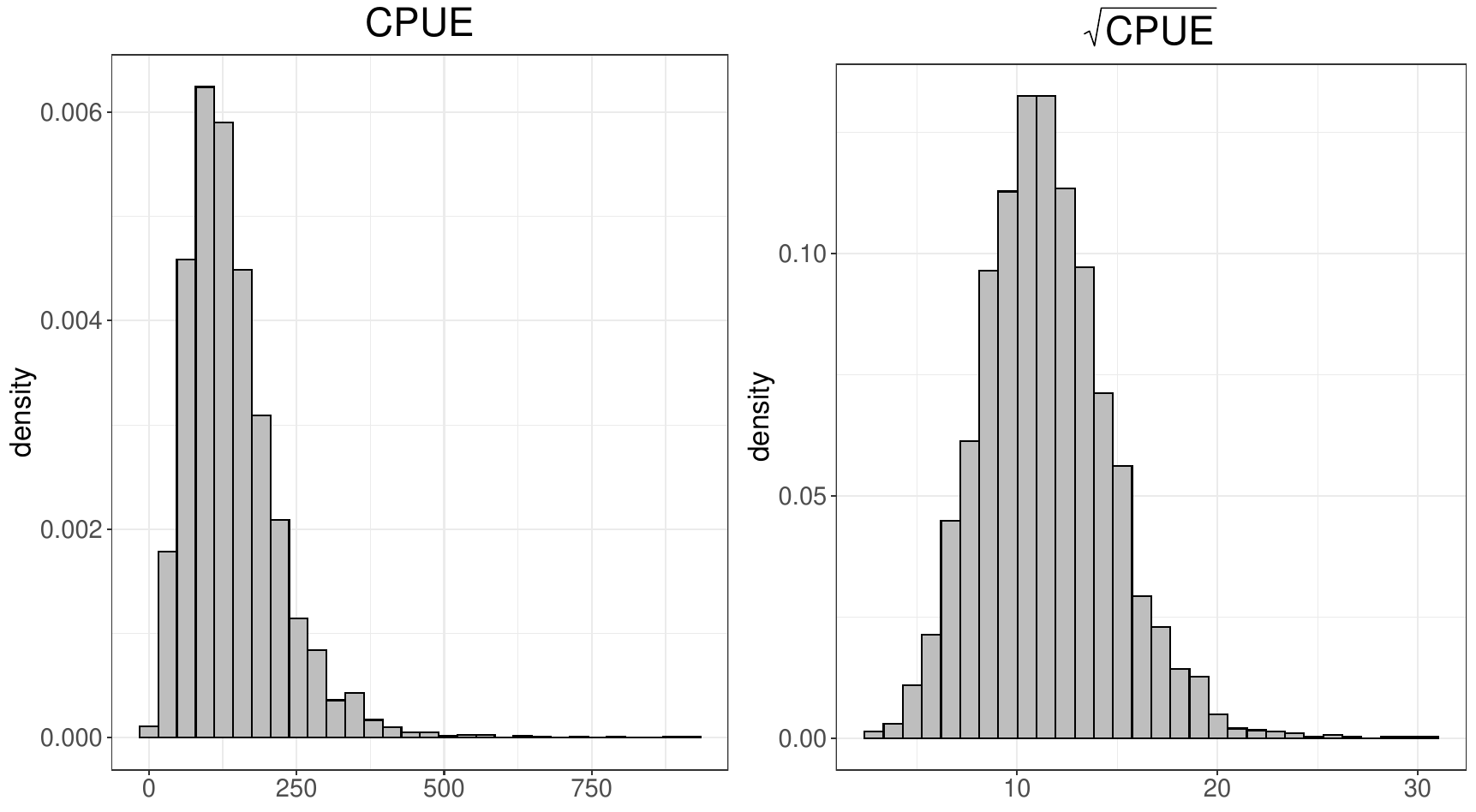}
\caption{Distributions of catch per united effort (CPUE) of Chilean sea urchin(left) and CPUE transformed via square root (right).}\label{fig:fig5}
\end{figure}

The analysis is based on the results presented in \cite{cavieres2021accounting}, where the authors introduced a Bayesian gamma spatio-temporal model utilizing a $\approx$GRF through the SPDE method. The new proposal is based in the results presented by \cite{cavieres2023thin}, where the authors developed a semiparametric spatio-temporal model using a TPS as spatial random effect. In this case, $\epsilon_{i}$ has a Skew normal distribution ($\epsilon_{i} \sim \mathcal{SN}(\mu_{i}, \sigma, \omega)$), and therefore $y_{i}$ as well. Given the above, we will adapt that model to a hierarchical structure enabling its utilization for Bayesian inference. Figure \ref{fig:fig5} shows the distribution of the CPUE and $\sqrt{\text{CPUE}}$ which will be used in the gamma and skew normal models respectively.

\newpage
~\newpage
Again, for comparative purposes, we rename the models as:

\begin{enumerate}
    \item \textbf{M1} $\rightarrow$ Gamma likelihood + $\approx$GRF as spatial random effect.
    \item \textbf{M2} $\rightarrow$ Skew normal likelihood + $\approx$GRF as spatial random effect.
    \item \textbf{M3} $\rightarrow$ Gamma likelihood + TPS as spatial random effect.
    \item \textbf{M4} $\rightarrow$ Skew normal likelihood + TPS as spatial random effect.
\end{enumerate}

Table \ref{table:table3} shows the structure of the models in the real data application for comparative purposes.

\begin{table}[h!]
\centering
\small{
\caption{Structure of the models in the real application.}
\begin{tabular}{ccc}
\hline
Model        & Structure                                   & Priors\\
\hline
             &                          &            \\
\textbf{M1}  &  $\text{\textbf{CPUE}} \mid \boldsymbol{u}, \boldsymbol{\theta} \sim  \text{Gamma}(\mu, \mu^{2}/\phi)$     &       $\boldsymbol{\beta} \sim \mathcal{N}(0, 5)$ \\
             &   $\eta_i = \log(\mu_i) = \boldsymbol{x}_{i}^{T}\boldsymbol{\beta} + u(\boldsymbol{s}_{i})$    &  $\tau \sim \mathcal{N}(0, 1)$    \\
             &   $\boldsymbol{u} \sim  \mathcal{N}(\boldsymbol{0}, \boldsymbol{Q}(\boldsymbol{\theta})^{-1})$ &  $\kappa \sim \mathcal{N}(0, 1)$  \\
             &   $\boldsymbol{\theta} = \{\text{log}(\tau), \text{log}(\kappa)\}$                             &                                   \\
\addlinespace
\addlinespace
\addlinespace
\addlinespace
\textbf{M2}  &  $\sqrt{\textbf{CPUE}}  \mid \boldsymbol{u}, \boldsymbol{\theta} \sim \text{Skew normal}(\boldsymbol{\mu}, \sigma, \omega)$  & $\boldsymbol{\beta} \sim \mathcal{N}(0, 5)$ \\ 
             & $ \mu_{i} = \boldsymbol{x}_{i}^{T}\boldsymbol{\beta} + u(\boldsymbol{s}_{i})$ & $\omega \sim \mathcal{N}(0, 1)$ (slant parameter)\\
             &   $\boldsymbol{u} \sim  \mathcal{N}(\boldsymbol{0}, \boldsymbol{Q}(\boldsymbol{\theta})^{-1})$  & $\sigma \sim \text{Cauchy}(0, 2)$                \\
             &   $\boldsymbol{\theta} = \{\text{log}(\tau), \text{log}(\kappa)\}$            & $\tau \sim \mathcal{N}(0, 1)$\\
             &                                                                               & $\kappa \sim \mathcal{N}(0, 1)$\\
\addlinespace
\addlinespace
\addlinespace
\addlinespace
\textbf{M3}  &  $\text{\textbf{CPUE}} \mid \boldsymbol{f}, \boldsymbol{\theta} \sim  \text{Gamma}(\mu, \mu^{2}/\phi)$     &       $\boldsymbol{\beta} \sim \mathcal{N}(0, 5)$                                                                                                       \\
             &   $\eta_i = \log(\mu_i) = \boldsymbol{x}_{i}^{T}\boldsymbol{\beta} + f(\boldsymbol{s}_{i})$   &  $\boldsymbol{c} \sim \mathcal{N}(0, 1)$\\
             &   $\boldsymbol{f} = \boldsymbol{E}_{k} \boldsymbol{c} $                                           &     \\
\addlinespace
\addlinespace
\addlinespace
\addlinespace
\textbf{M4}  &  $\sqrt{\textbf{CPUE}} \mid \boldsymbol{f}, \boldsymbol{\theta} \sim \text{Skew normal}(\boldsymbol{\mu}, \sigma, \omega)$   & $\boldsymbol{\beta} \sim \mathcal{N}(0, 5)$      \\ 
             &  $ \mu_{i} = \boldsymbol{x}_{i}^{T}\boldsymbol{\beta} + f(\boldsymbol{s}_{i})$      & $\omega \sim \mathcal{N}(0, 1)$ (slant parameter)\\
             &   $\boldsymbol{f} = \boldsymbol{E}_{k} \boldsymbol{c} $                             & $\sigma \sim \text{Cauchy}(0, 2)$                \\
             &                                                                                     & $\lambda \sim \mathcal{N}(0, 1)$ (penalty parameter)\\
             &                                                                                     & $\boldsymbol{c} \sim \mathcal{N}(0, 1)$ \\
\addlinespace
\addlinespace
\addlinespace
\hline
\end{tabular}
\label{table:table3}}
\end{table}
\footnotesize{In the Gamma likelihood function, $\phi$ is a precision parameter}.
\normalsize

\pagebreak 
\subsection{Real case study results}

\begin{figure}[hbt!]
\centering
\includegraphics[width=15.5cm, height=9cm]{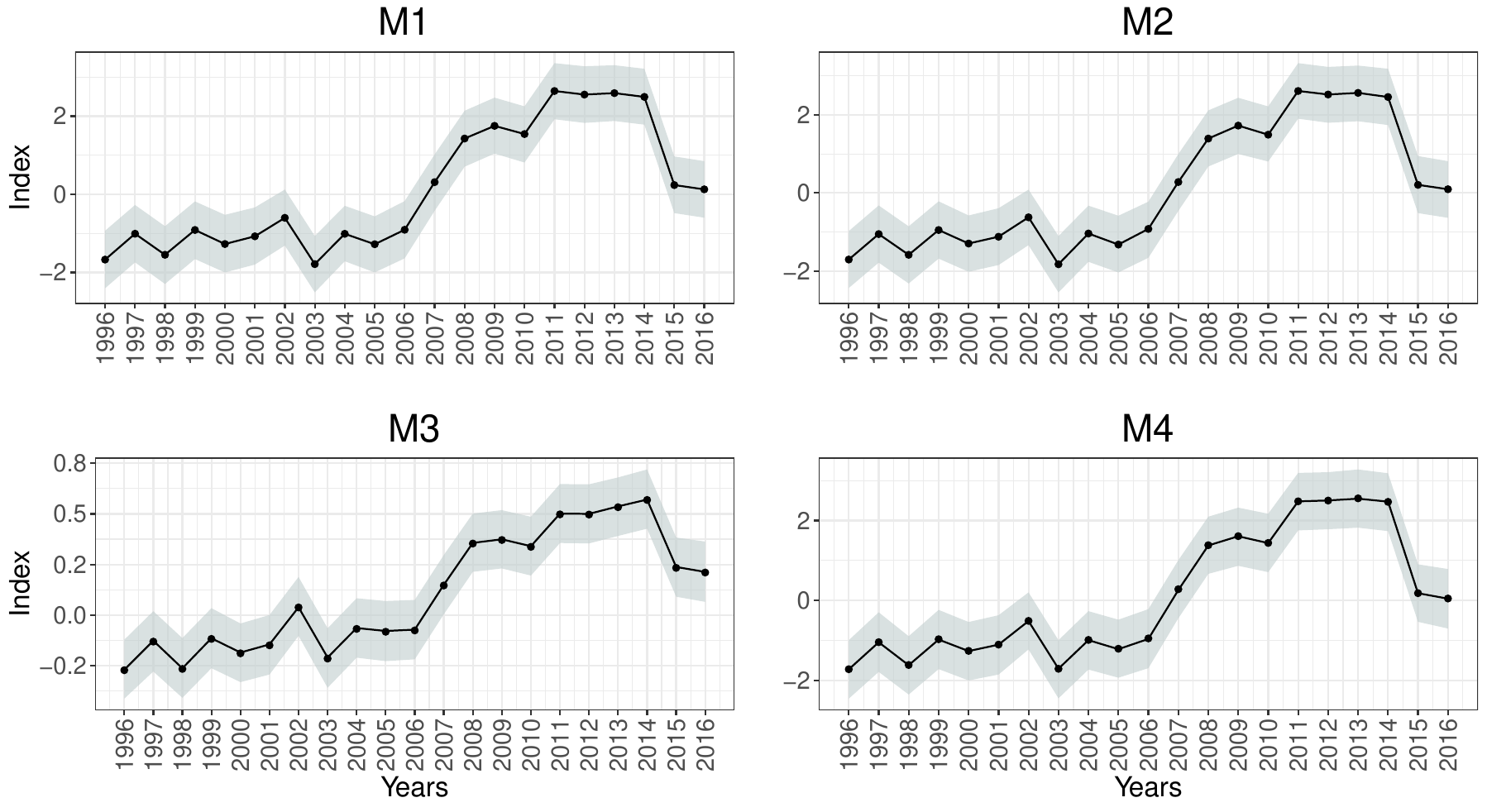}
\caption{Relative index of abundance estimated by the \textbf{M1}, \textbf{M2}, \textbf{M3} and \textbf{M4}.}\label{fig:fig6}
\end{figure}

Figure \ref{fig:fig6} shows the relative index of abundance for the sea urchin (\textit{Loxechinus albus}) from 1966 to 2016 calculated with \textbf{M1}, \textbf{M2}, \textbf{M3}, and \textbf{M4}. The models generally estimate similar trends and uncertainty across time, but there are few differences between the estimated indexes. For example, in 2002, the index for \textbf{M1} is close to around -0.5, whereas for \textbf{M4} in the same year, it is approximately -0.2. Another small difference was found in the estimates from 2015 to 2016. where the decline of \textbf{M4} is slightly more pronounced, but almost imperceptible when considering the general trend of all the models. The only model exhibiting distinct behavior is \textbf{M3}, where the decline from 2014 to 2015 appears to be less pronounced, and the absolute scale of the index is notably smaller from that of other models (with the posterior mean ranging from -0.2 to 0.7).   

\begin{figure}[hbt!]
\centering
\includegraphics[width=16cm, height=6.5cm]{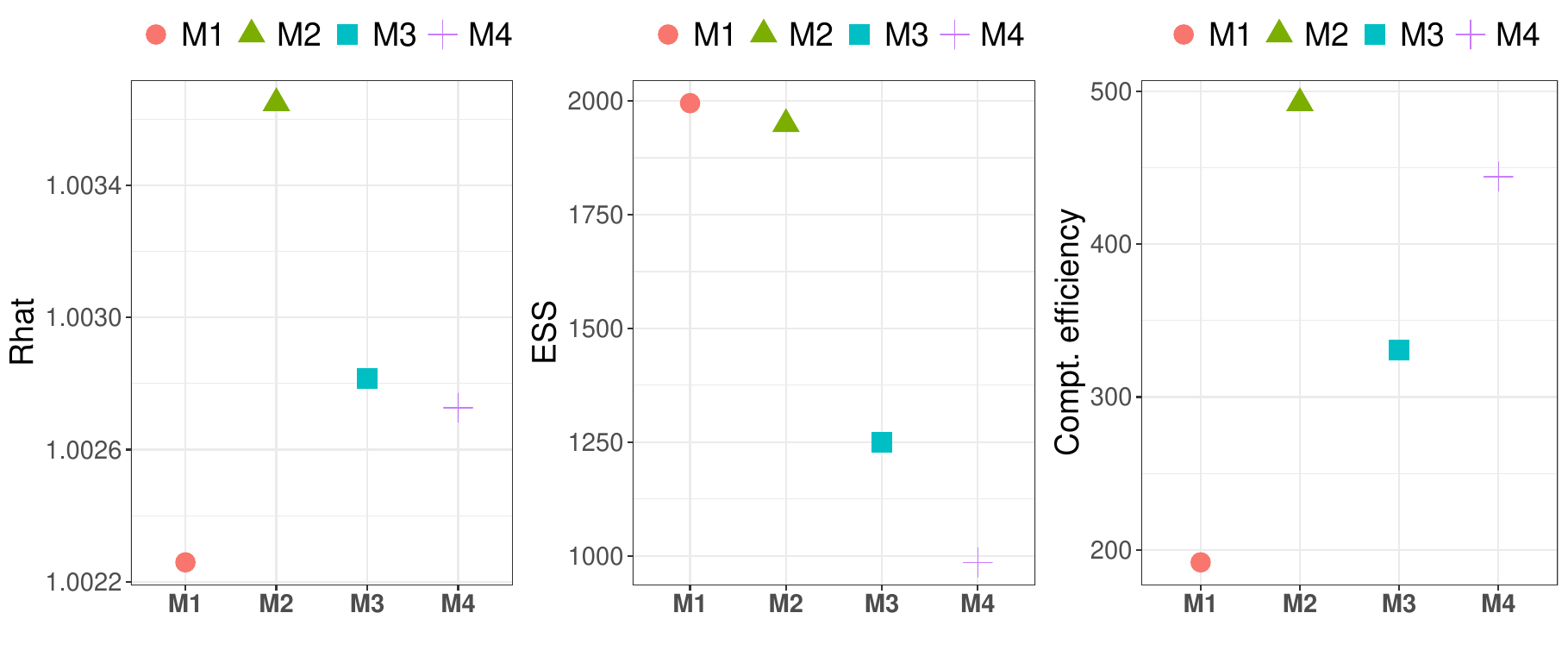}
\caption{Rhat, ESS and the computational efficiency for \textbf{M1}, \textbf{M2}, \textbf{M3} and \textbf{M4}.}\label{fig:fig7}
\end{figure}

Since no models had divergences in the simulated trajectories of the HMC (\ref{appendix:appendix2}),  meaning that simulated Hamiltonian trajectory are reliable (\cite{stan2018stan})), and the models present similar estimation for the indexes, we also compare the Rhat, the ESS and the computational efficiency. As  Figure \ref{fig:fig7} shows, the values of the Rhat is in the range of 1.002 (\textbf{M1}) and 1.005 (\textbf{M3}), indicating that the four models show no evidence of lack of convergence based on the criteria Rhat $<$ 1.05 established by \cite{aki2020rank}. In relation to the ESS, and as this value should be as large possible, the models with a high number of effective samples are \textbf{M1} and \textbf{M2}. However, based on other diagnostics, the posterior distribution of the parameters remained consistent for both \textbf{M3} and \textbf{M4}. Finally, the models \textbf{M2} and \textbf{M4} presented the best computational efficiency respectively (Figure \ref{fig:fig7}).  

To identify the model with the the best fit to the data, while balancing parsimony and overfitting, we used leave-one-out cross-validation (loo-cv), specifically Pareto smoothed importance sampling (PSIS-LOO, \cite{vehtari2017practical}). PSIS-LOO is used to compute components of el\_pd (expected log predictive density), and in turn will be used for model comparison. PSIS-LOO provides a diagnostic metric $\hat{k}$ such that when  $\hat{k} < \text{min}(1 - 1/\text{log10}(\text{S}), 0.7)$ then the model is well specified, otherwise the model could indicate problems (e.g. influence of extreme values in the data). In the previous expression, S is the sample size. It is important to point out that in the comparison we also included the Jacobian correction for the \textbf{M2} and \textbf{M4} models, since the response variable, CPUE, was transformed to $\sqrt{\text{CPUE}}$ in the Skew normal models. 

\begin{figure}[hbt!]
\centering
\includegraphics[width=15cm, height=8cm]{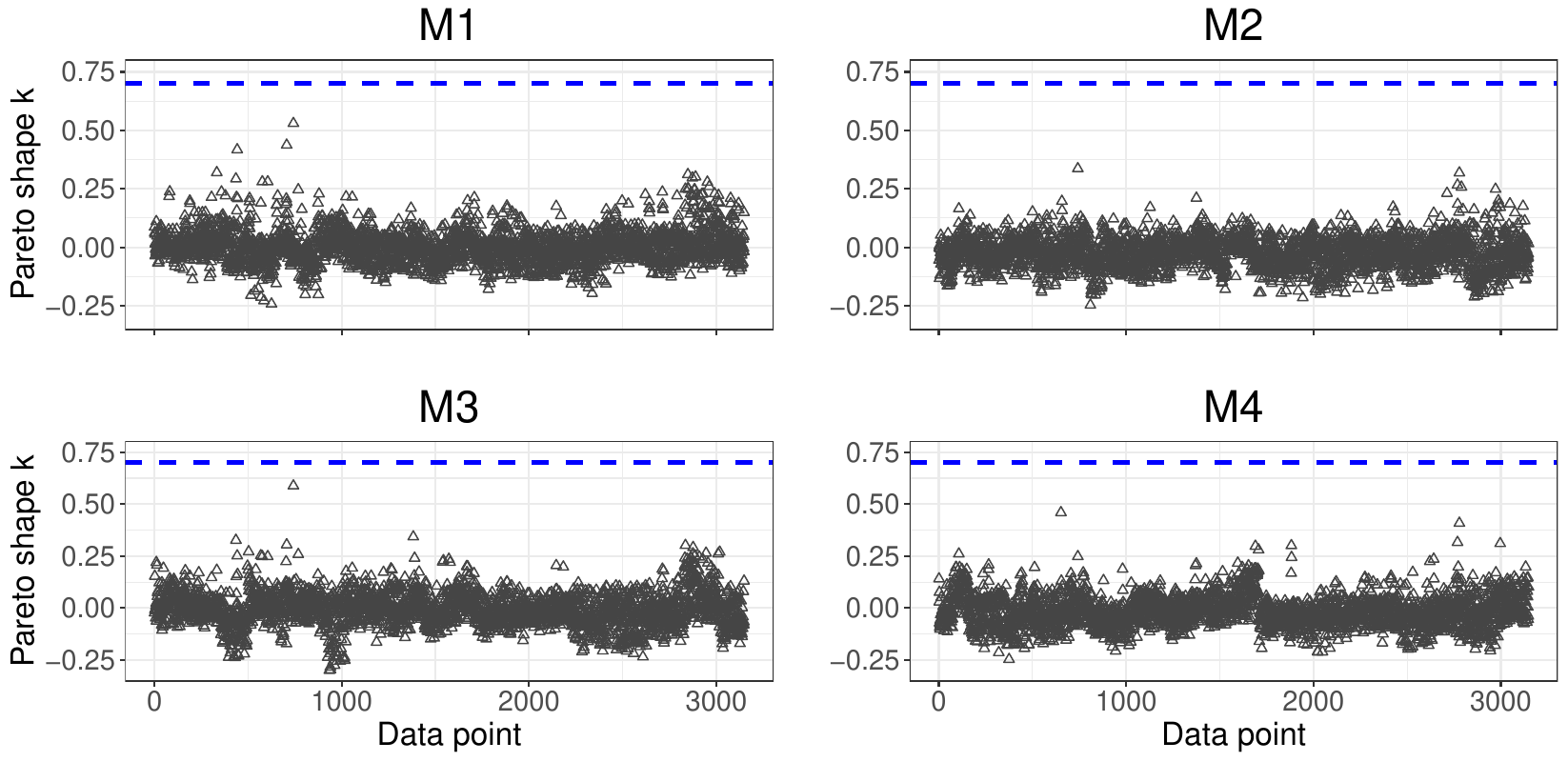}
\caption{Pareto shape $k$ contribution for the comparative analysis. In the four models, $k < 0.71$ (segmented blue line), signifying a good psis\_loo approximation.}\label{fig:fig8}
\end{figure}

From Figure \ref{fig:fig8} we can see that \textbf{M1} and \textbf{M3} exhibits one value between 0.5 - 0.7, while the $\hat{k}$ values for \textbf{M2} and \textbf{M4} are less than 0.5, indicating a reliable estimation of el\_pd. The Figure \ref{fig:fig8} shows the $\hat{k}$ values for the four models and the threshold of 0.71 assumed as diagnostic to model evaluation considering the size of the sample.    

\textbf{M1} and \textbf{M2} have more parameters than \textbf{M3} and \textbf{M4}, specifically they contain 163 parameters in total, 27 associated with the linear fixed effects, 3 related to hyperparameters ($\sigma$, $\tau$ and $\kappa$), and 131 associated to the spatial random effect. The parameters of the $\approx$GRF are entirely determined by the number of triangles in the discretization of the continuous spatial domain. In contrast, \textbf{M3} and \textbf{M4} contains the same number of parameters for fixed effects (27), 3 hyperparameters ($\sigma$, $\omega$ and $\lambda$), and only 12 parameters related to the TPS (the truncated basis functions). The difference in the number of parameters is due to the difficulty in decreasing the number of triangles in the mesh (to build the discretized spatial domain) where the $\approx$GRF is introduced as spatial random effect in \textbf{M1} and \textbf{M2}. For a smaller number of triangles in the discretized spatial domain, the convergence in the chains would have been impossible. 

Although the effective number of parameters in the models are different, they are much less than the number of observations. Therefore, the specification of the models should not be a problem for comparative purposes. 
The comparison using the elpd\_diff criterion show us that the \textbf{M1} model has a better predictive performance than \textbf{M2}, \textbf{M3} and \textbf{M4}. That is, \textbf{M1} has a better expected predictive accuracy as determined by the differences in expected log-predictive distribution (elpd\_loo) (Table \ref{table:table4}).  

\begin{table}[hbt!]
\centering
\caption{loo-cv criterion for comparative purposes. p\_loo is assumed as the effective number of parameters, elpd\_diff measures the difference between each model relative to the best $\widehat{\text{elpd}}$ (the model in first row), se\_diff is the standard error of the difference in elpd\_diff and the execution time of estimation for each model.}
\label{table:table4}
\begin{tabular}{ccccc}
\hline
 & p\_loo &  elpd\_diff & se\_diff & Execution time\\
\hline
\textbf{M1}     &  45.16 &  0.00   &  0.00   & 10.40 hrs\\
\textbf{M3}     &  40.76 &  -6.27  &  5.76   & 3.78 hrs\\
\textbf{M4}     &  37.82 &  -51.04 &  25.25  & 2.22 hrs\\
\textbf{M2}     &  27.86 &  -93.11 &  29.16  & 3.31 hrs\\
\hline
\end{tabular}
\end{table}

\begin{figure}[hbt!]
\centering
\includegraphics[width=16cm, height=6.5cm]{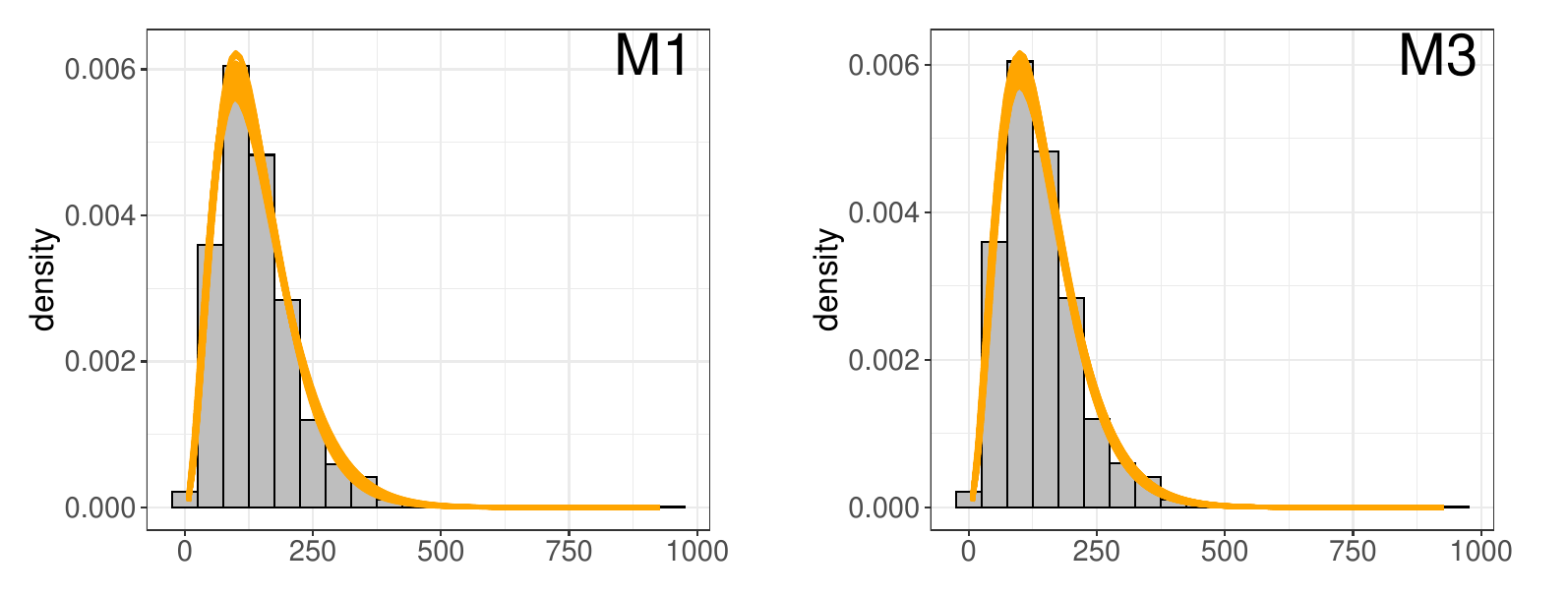}
\caption{Simulated (200) posterior predictive distributions from \textbf{M1} and \textbf{M3} (yellow lines)}\label{fig:fig9}
\end{figure}

Although \textbf{M1} has a better predictive performance, the difference between this model and $\textbf{M3}$ (the second best predictive model) are minimal (Figure \ref{fig:fig9} illustrates that both models accurately represent the distribution of the CPUE data based on their posterior predictive distributions). Besides, the execution time of \textbf{M1} is 2.75 time higher than \textbf{M3} (Table \ref{table:table4}) and their 

Nevertheless, as mentioned earlier, setting the correct prior distributions for the hyperparameters in \textbf{M1} and \textbf{M2} proves challenging for the sea urchin data, particularly when we are interested in having a Bayesian model that satisfy convergence criteria and their diagnostics. Since this process could be time consuming, given the need for the construction of a correct mesh and setting appropriate priors on the hyperparameters, the execution time required for the estimation process of \textbf{M3} and \textbf{M4} (the second and third best predictive models) are not the only advantage to consider, but its simple implementation as well.

\section{Conclusions}
\label{section6}

Using a TPS as spatial random effect for spatial modelling purposes presents a valid alternative to the approximated GRF for Bayesian inference, specifically for an MCMC method. While it is not commonly employed in this framework, we demonstrated that it enables fast computation when considering a low-rank approximation of the spline matrices. Furthermore, it is simple to parameterize and provides an accurate estimation of parameters in the model, similar performance to a model using an approximated GRF, an order of magnitude faster for the mesh sizes and truncation considered in the analyses.

In the first analysis of this study, the Bayesian spatial model which incorporates a TPS (\textbf{M-TPS}) demonstrated high performance in recovering the values for the parameters used in the simulation, where these values were similar to those estimated by \textbf{M-GRF}. Furthermore, the execution time in each simulated spatial locations was less as well. In this same analysis, the posterior predictive distribution derived of the parameters estimated by \textbf{M-GRF} and \textbf{M-TPS} presents similar behaviour to represent the simulated response variable. In terms of computational efficiency (min(ESS)/time), \textbf{M-TPS} outperformed \textbf{M-GRF} for all simulated models across different sizes of spatial locations. 

In the second analysis, which involved a real application, the four models (\textbf{M1}, \textbf{M2}, \textbf{M3} and \textbf{M4}) displayed similar trends in the estimated index, along with their respective uncertainty (based on the 25\textit{th} and 75\textit{th} percentiles of the posterior distribution for the parameters). We note that \textbf{M3} had a different absolute scale (Figure \ref{fig:fig6}), but because these models are measuring relative abundance and the stock assessment includes an estimated "catchability" scaling term we do not consider this an immediate issue given the focus on estimation performance. However, further investigation into the cause of this absolute scale discrepancy would be prudent in a real analysis. 
In this case, and based on the loo-cv criterion, \textbf{M1} exhibits superior predictive performance compared to the others. However, the difference in the expected log-predictive density is minimal compared with \textbf{M3} for a specific number of predictive simulations. Additionally, the execution time needed to fit \textbf{M1} was 2.75 times that required for fitting \textbf{M3}, reflecting a notable advantage in favor of \textbf{M3} if we consider the computational efficiency of this model. In the computational efficiency comparison, \textbf{M2} demonstrated the best behaviour. However, its predictive performance was the lowest among all the models. 

The main advantage of the TPS as spatial random effect is its straightforward parametrization since is not necessary to estimate parameters that govern the relationship between sites (locations). Hence, it is not necessary to impose any form of stationarity as is assumed when we use an approximated GRF as spatial random effect. On the other hand, a clear disadvantage emerges from the absence of parameters within the RBFs, rendering us unable to characterize the underlying spatial process. For example, the GRF approach allows for statements about decorrelation distance and anisotropy which may be valuable in certain contexts. However, TPS replaces the process of mesh construction with an algorithm that efficiently truncates basis functions, which is a sensitive part to build the approximated GRF. Furthermore, in conventional MCMC algorithms, the parameters of the covariance function might exhibit correlations that impede the proper convergence of the chains. An alternative approach to dealing this issue may be to use Penalized Complexity priors (PC priors) for the range ($\rho$) and marginal variance ($\sigma_{u}$) parameters. This parameterization could avoid dependency issues in the hyperparameters of the spatial random field (such as those commonly happens with $\tau$ and $\kappa$).




\newpage
\bibliographystyle{unsrtnat}
\bibliography{references}

\vskip1cm

\newpage
\appendix
\section{\textbf{M-TPS} computational efficiency}\label{appendix:appendix1}

Figure \ref{fig:fig1} could suggest that increasing the number of spatial locations (SL) may lead to similar computational efficiency for \textbf{M-GRF} and \textbf{M-TPS}, however, this idea is not entirely true. While increasing the number of knots for the TPS may indeed lead to comparable computational efficiency between \textbf{M-GRF} and \textbf{M-TPS}, the increase of the number of knots is not necessary at all. As we know that execution computation of \textbf{M-TPS} is faster than \textbf{M-GRF}, we will do the comparison  using the same model, namely; 

\begin{itemize}
    \item Fitting a model where the number of knots is increasing as SL does $\rightarrow$ \textbf{M-TPS}
    \item Fitting a model where the number of knots is fixed but the SL increases $\rightarrow$ \textbf{M-TPS(2)}
\end{itemize}

Let's consider SL ranging from 700 to 1000, denoting them as SL7, SL8, SL9, and SL10, respectively. For the above, we compare the \textbf{M-TPS} with \textbf{M-TPS(2)}, which represents the same model but with 30 fixed knots. 
First, we will focus in the parameters estimated by \textbf{M-TPS} and \textbf{M-TPS(2)} (Table \ref{table:tableapp1}). We observe that the estimated parameters are very similar between the models, with values closer to those used in the simulated data. However, there is a huge difference in the execution time, where \textbf{M-TPS} may take until 8 times more than the execution time of \textbf{M-TPS(2)} (Table \ref{table:tableapp1}).  

\begin{table}[!htbp]
\setlength{\tabcolsep}{4pt}
\caption{Means, $25\%$, and $75\%$ percentiles of the posterior distributions, relative error and execution time for \textbf{M-TPS} and \textbf{M-TPS(2)} in the different simulated spatial locations (SL).}
\centering
\scalebox{0.95}{
\begin{tabular}{cccccccccccc}
\hline
    &  & \multicolumn{5}{c}{\textbf{M-TPS}} &   \multicolumn{5}{c}{\textbf{M-TPS(2)}}\\
   \hline
 SL   & Real values & Mean & 25\% & 75\% & R.error & E.time     & Mean  & 25\% & 75\% & R.error & E.time \\
 \hline
\multirow{3}{*}{SL7} & $\beta_{0} = 1.0$  & 1.03 & 1.01 & 1.05 & 0.03 & \multirow{3}{*}{\textcolor{red}{$\sim$ 42 min}} & 1.00 & 1.00 & 1.01 & 0.00&  \multirow{3}{*}{\textcolor{blue}{$\sim$ 6 min}} \\
                     & $\beta_{1} = 2.0$  & 1.97 & 1.94 & 2.00 & 0.03 &        & 1.99 & 1.99 & 2.00 & 0.01\\
                     & $\sigma    = 0.1$  & 0.11 & 0.11 & 0.12 & 0.01 &        & 0.10 & 0.10 & 0.10 & 0.00\\
\hline
\multirow{3}{*}{SL8}& $\beta_{0}  = 1.0$ & 1.00  & 1.00 & 1.01 & 0.00 & \multirow{3}{*}{\textcolor{red}{$\sim$ 58 min}} & 1.00 & 0.99 & 1.00 & 0.00 &  \multirow{3}{*}{\textcolor{blue}{$\sim$ 10 min}} \\
                    & $\beta_{1}  = 2.0$ & 2.00  & 1.99 & 2.01 & 0.01 &        & 2.00 & 2.00 & 2.01 & 0.00\\
                    & $\sigma     = 0.1$ & 0.10  & 0.10 & 0.10 & 0.00 &        & 0.10 & 0.10 & 0.10 & 0.00\\
\hline
\multirow{3}{*}{SL9}& $\beta_{0}  = 1.0$  & 1.00 & 1.00 & 1.01 & 0.00 & \multirow{3}{*}{\textcolor{red}{$\sim$ 63 min}} & 1.00 & 0.99 & 1.00 & 0.00 & \multirow{3}{*}{\textcolor{blue}{$\sim$ 11 min}} \\
                    & $\beta_{1}  = 2.0$  & 1.99 & 1.99 & 2.00 & 0.01 &        & 2.01 & 2.00 & 2.01 & 0.01  \\
                    & $\sigma     = 0.1$  & 0.10 & 0.10 & 0.10 & 0.00 &        & 0.10 & 0.10 & 0.11 & 0.00 \\
    \hline
\multirow{3}{*}{SL10}& $\beta_{0} = 1.0$  & 1.00 & 1.00  & 1.01 & 0.00 & \multirow{3}{*}{\textcolor{red}{$\sim$ 99 min}} & 1.00 &0.99 & 1.01 &0.00  & \multirow{3}{*}{\textcolor{blue}{$\sim$ 12 min}} \\
                     & $\beta_{1} = 2.0$  & 2.00  & 1.99 & 2.01 & 0.00 &       & 2.00 & 1.99 & 2.00 & 0.00\\
                     & $\sigma    = 0.1$  & 0.10  & 0.10 & 0.11 & 0.00 &       & 0.10 & 0.10 & 0.11 & 0.00\\
\hline
\end{tabular}}
\label{table:tableapp1}
\end{table}
\footnotesize{R.error = $||\text{Real value} - \text{Mean}||$
\vspace{0.2cm}

E.time = Execution time.}
\normalsize

\begin{figure}[hbt!]
\centering
\includegraphics[width=11.5cm, height=7.5cm]{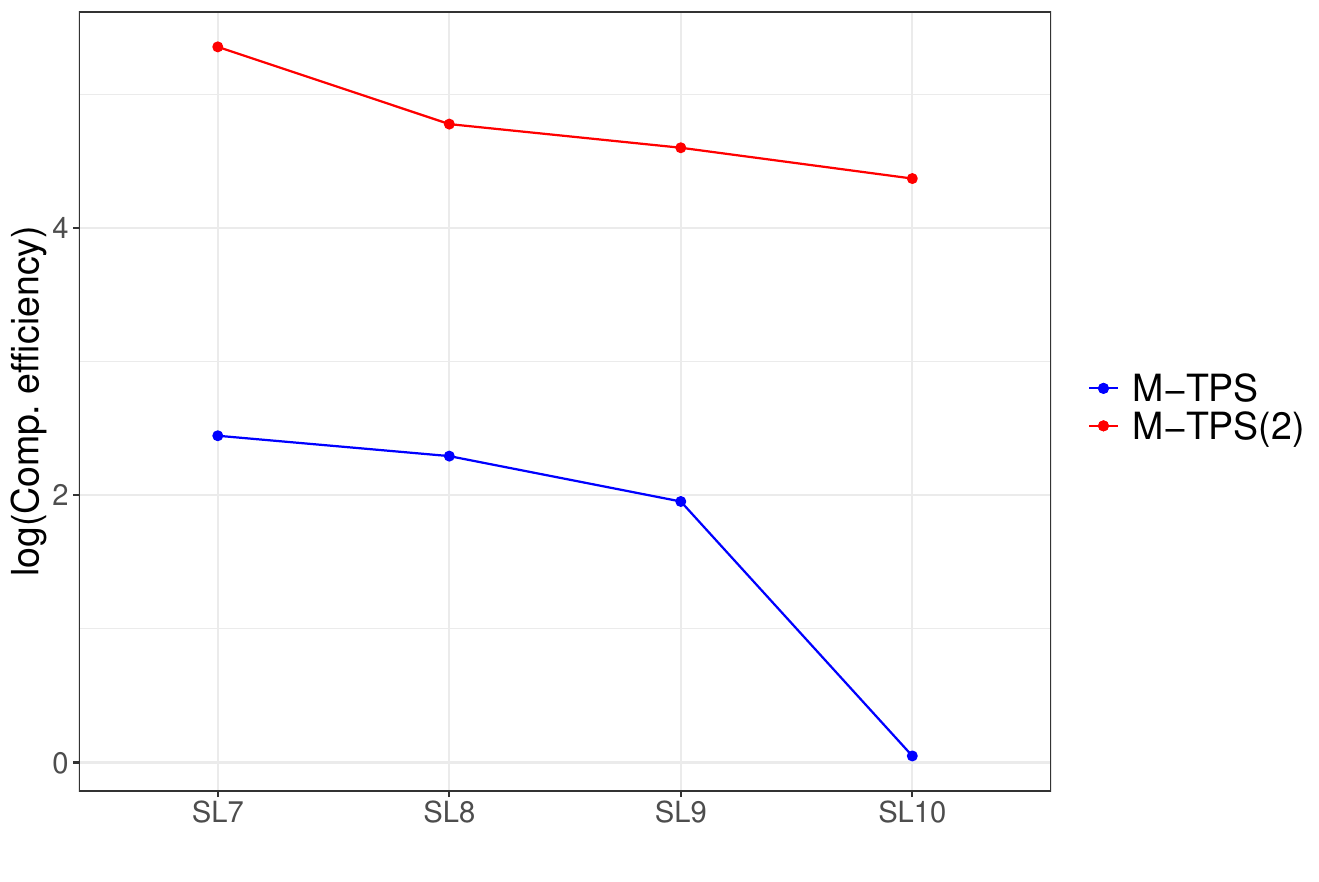}
 \caption{Computational efficiency in log scale (log(\texttt{min (ESS)/ time})) calculated for different number of spatial locations (SL) in \textbf{M-TPS}.}
 \label{fig:fig10}
\end{figure}

\newpage
\normalsize 
Figure \ref{fig:fig10} shows the computational efficiency for both models. It is clear that \textbf{M-TPS(2)} is much more efficient than \textbf{M-TPS}, hence a small number of knots to approximate the TPS allows us obtain a good performance of the model considering the values estimated for the parameters and, especially, in the execution time.

\begin{figure}[hbt!]
\centering
\includegraphics[width=16cm, height=8cm]{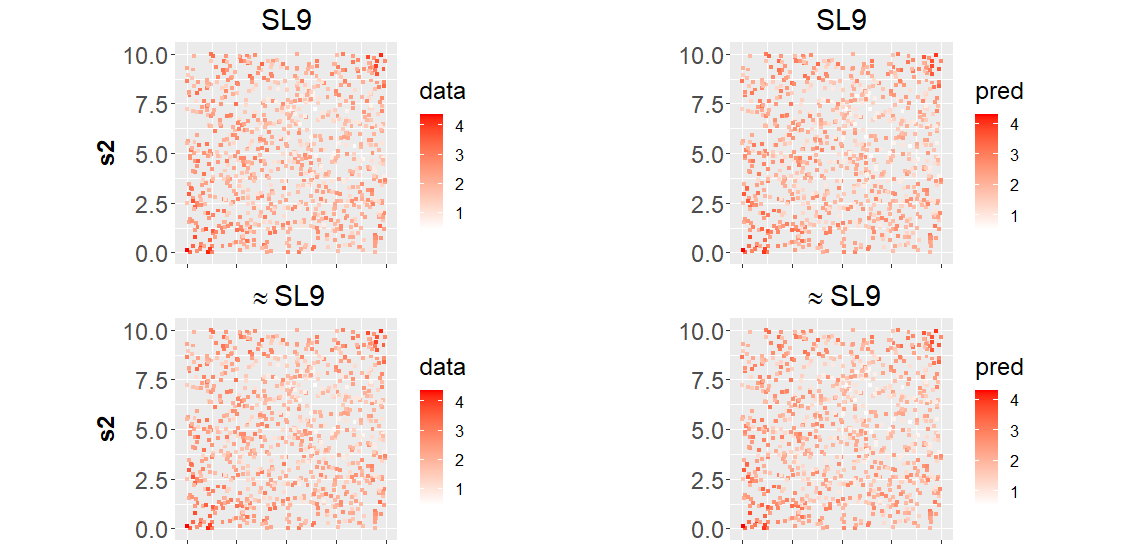}
 \caption{Prediction of the simulated response variable ``y\_sim" for SL9 (top panel) and $\approx$SL9 (bottom panel). Both models are using the \textbf{M-TPS} structure to fit the data.}
 \label{fig:fig11}
\end{figure}

Finally, we evaluate the spatial predictions of two models; one model with a large number of knots and another model with a small number. For this, let's consider SL9 (900 random spatial points) which contains 90 knots, and another model $\approx$L9 containing 30 knots. Figure \ref{fig:fig11} demonstrates that the simulated response variable, ``data" is accurately predicted by SL9 (top panel). This accurate prediction is given the estimation of a suitable number of basis functions, enabling us to effectively represent the underlying spatial process of the model. Conversely, $\approx$SL9 also correctly predicts ``data" despite the smaller number of knots in the model. Thus, using the truncated basis functions of the TPS allows us properly obtain the parameters used for the simulated data and also predict the response variable into spatial domain with a high accuracy (Table \ref{table:tableapp2}). 

\begin{table}[!htbp]
\caption{Statistical performance of spatial prediction for SL9 and $\approx$SL9.}
\centering
\begin{tabular}{ccc}
\hline
SL & number of knots & RSME\\
\hline
SL9          & 90  & 2.55e-05 \\
$\approx$SL9 & 30  & 9.79e-05 \\
 \hline
\end{tabular}
\label{table:tableapp2}
\end{table}

\newpage
\section{Additional diagnostic plots}\label{appendix:appendix2}


\begin{figure}[hbt!]
\centering
\includegraphics[width=13cm, height=7cm]{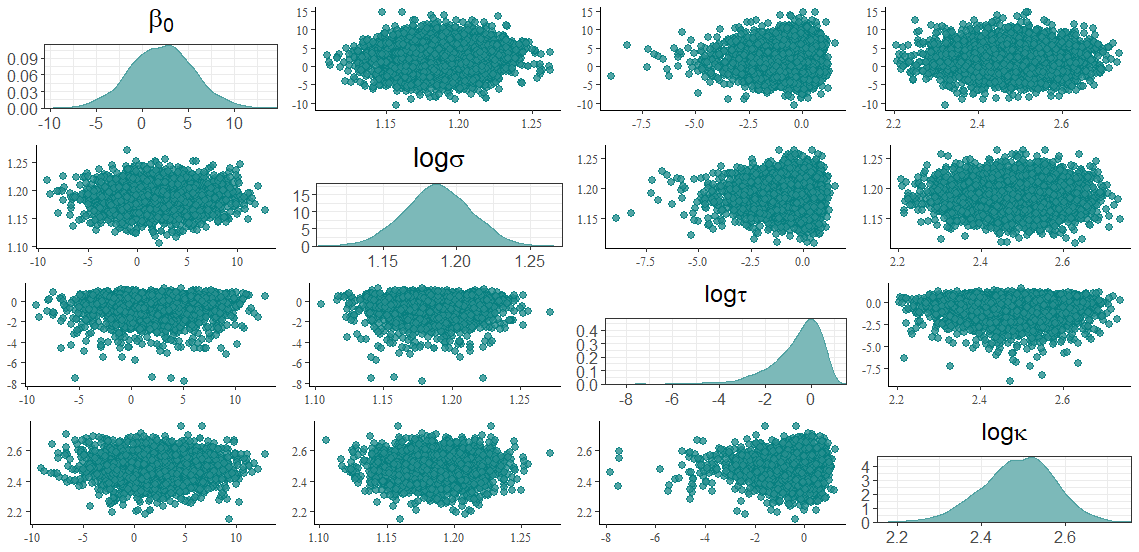}
\caption{Pairs plot for the parameter $\beta_{0}$ and hyperparameters of the spatial effect $\text{log}(\sigma)$, $\text{log}(\tau)$, $\text{log}(\kappa)$ in \textbf{M1}.}\label{fig:fig12}
\end{figure}

\begin{figure}[hbt!]
\centering
\includegraphics[width=13cm, height=7cm]{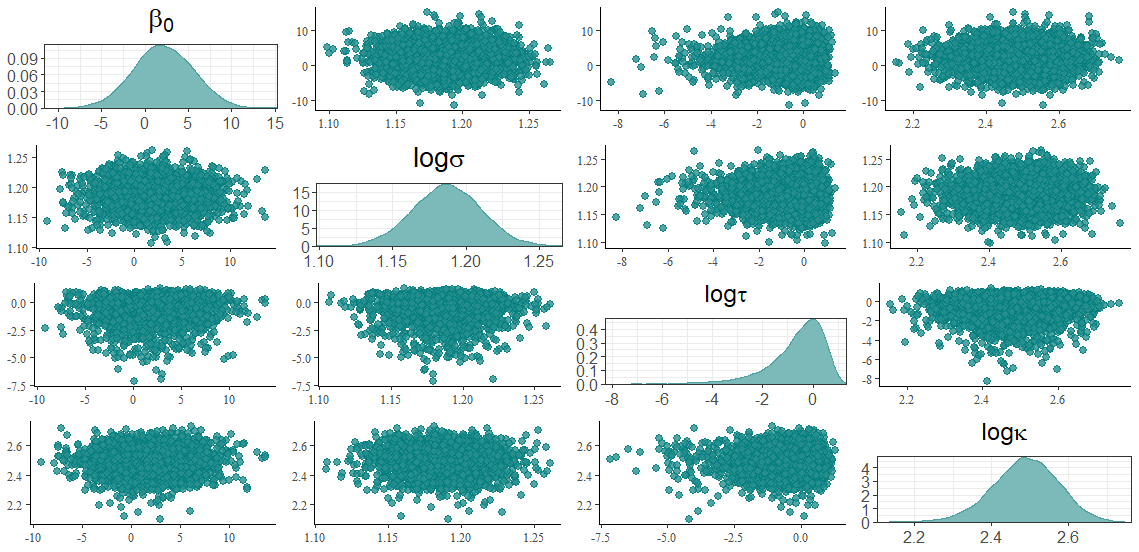}
\caption{Pairs plot for the parameter $\beta_{0}$ and hyperparameters of the spatial effect $\text{log}(\sigma)$, $\text{log}(\omega)$, $\text{log}(\lambda)$ in \textbf{M2}.}\label{fig:fig13}
\end{figure}

\newpage

\begin{figure}[hbt!]
\centering
\includegraphics[width=13cm, height=7cm]{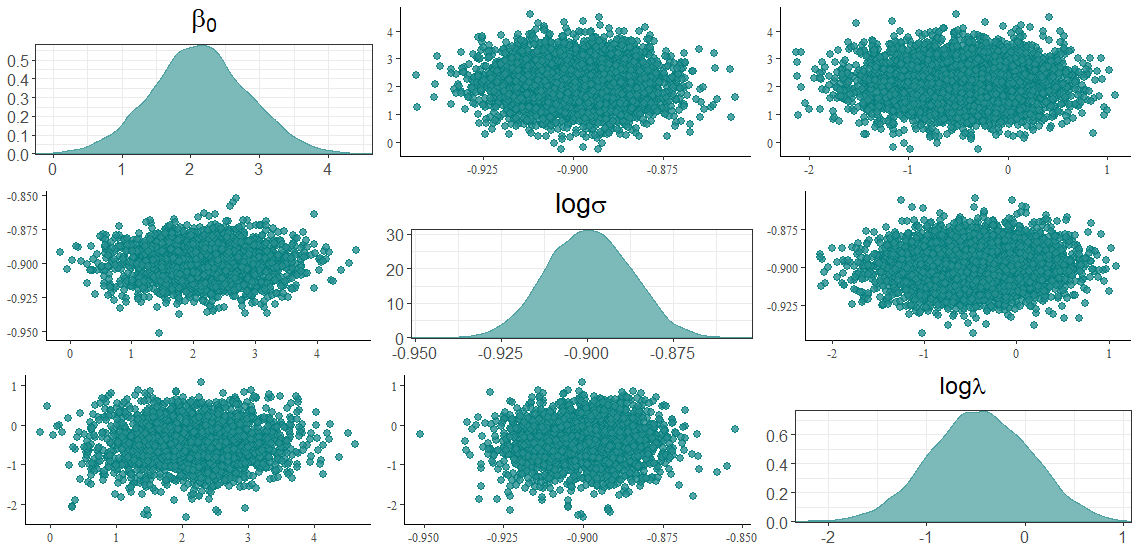}
\caption{Pairs plot for the parameter $\beta_{0}$ and hyperparameters of the spatial effect $\text{log}(\sigma)$, $\text{log}(\tau)$, $\text{log}(\kappa)$ in \textbf{M3}.}\label{fig:fig14}
\end{figure}

\begin{figure}[hbt!]
\centering
\includegraphics[width=13cm, height=7cm]{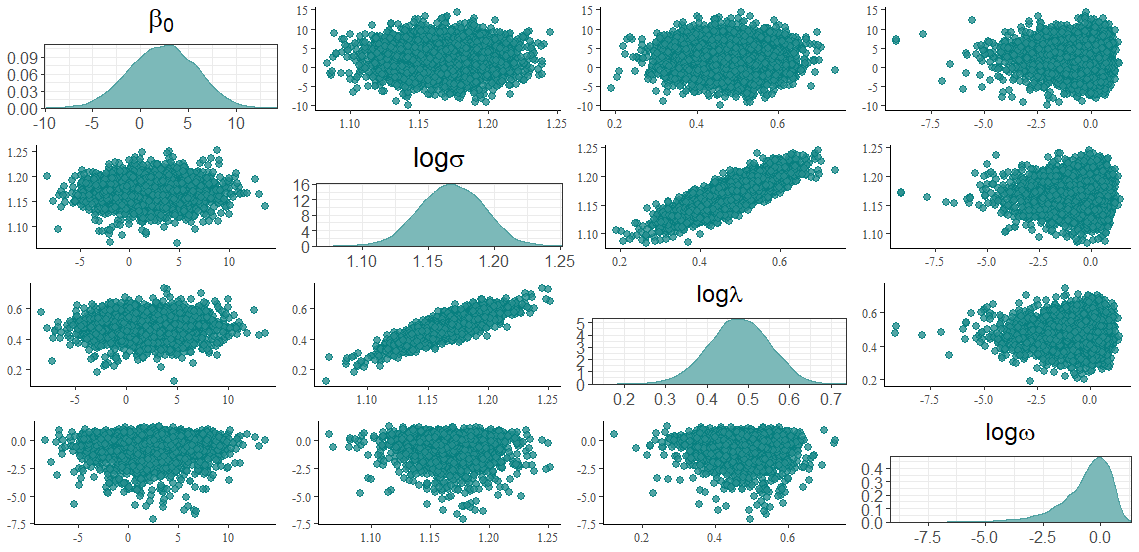}
\caption{Pairs plot for the parameter $\beta_{0}$ and hyperparameters of the spatial effect $\text{log}(\sigma)$, $\text{log}(\omega)$, $\text{log}(\lambda)$ in \textbf{M4}.}\label{fig:fig15}
\end{figure}




\end{document}